\documentclass[a4paper,11pt]{article}
\usepackage{thmtools,thm-restate}
\usepackage{umr-V2}
\usepackage[usenames,dvipsnames,pdftex]{color}
\usepackage{epic,eepic,amsmath,latexsym,amssymb}
\usepackage{ifthen,graphics,epsfig}
\usepackage[english]{babel}
\usepackage{times}
\usepackage{float}
\usepackage{xspace}
\usepackage[T1]{fontenc}
\usepackage{graphicx}

\usepackage[latin1]{inputenc}
\usepackage[T1]{fontenc}

\usepackage{MnSymbol}

\usepackage{amsmath}
\usepackage{fixltx2e}
\usepackage{graphicx}
\usepackage{longtable}
\usepackage{float}
\usepackage{wrapfig}
\usepackage{textcomp}
\usepackage{wasysym}
\usepackage{hyperref}
\usepackage{tikz}
\usepackage{varioref}
\usepackage{multirow}
\usepackage{rotating}
\usepackage{algorithmicx}
\usepackage{algpseudocode}
\usepackage{algorithm}

\algsetblockdefx{ForAllDo}{}
{0}{}
[2]{\textbf{for all} #1 \textbf{do} #2}

\algsetblockdefx{IfThenCommand}{}
{0}{}
[2]{\textbf{if} #1 \textbf{then} #2 \textbf{ end if}}

\algsetblockdefx{RepeatUntilOneLine}{}
{0}{}
[2]{\textbf{repeat} #1 \textbf{until} #2}

\algsetblockdefx{WaitUntil}{}
{0}{}
[1]{\textbf{wait until} #1}

\algdef{SnE}{ForAllDoBlock}{EndForAllDo}
[1]{\textbf{for all} #1 \textbf{do}}

\algdef{SE}{OnTocoDeliver}{EndTocoDeliver}
[1]{\textbf{on toco\_deliver} #1}{\textbf{end on toco\_deliver}}

\algdef{SE}{ProcessIs}{EndProcessIs}
[1]{\textbf{process} #1 \textbf{is}}{\textbf{end process}}

\algdef{SE}{OnReceive}{EndOnReceive}
[1]{\textbf{on receiving} #1}{\textbf{end on receiving}}

\algdef{SnE}{WaitUntilWhere}{EndWaitUntilWhere}
[1]{\textbf{wait until} #1 \textbf{where}}

\algdef{SE}{Background}{EndBackground}
[1]{\textbf{background task} #1 \textbf{is}}{\textbf{end background task}}

\algrenewcommand\algorithmicfunction{\textbf{operation}}

\newcommand{\func}[1]{\textsf{#1}}

\algrenewtext{Function}[2]%
  {\algorithmicfunction\ \func{#1}(#2)}

\renewcommand{\Call}[2]%
  {\textsf{#1}(#2)}

\algrenewcommand\algorithmicloop{\textbf{loop forever}}

\newcommand{\DoForAllEmbed}[2]{#1 \textbf{for all} #2}

\algsetblockdefx{DoForAll}{}
{0}{}
[2]{\DoForAllEmbed{#1}{#2}}

\newcommand{\skipSomeSpace}{\Statex \vspace{-0.5em}}

\tikzstyle{myRec}=[draw,rectangle,minimum height=0.22cm]
\tikzstyle{linearPoint}=[draw,rectangle,minimum height=0.32cm,minimum width=0.05cm,inner sep=0cm,fill=black]

\newcommand{\operation}[4]{%
  -- ({#1+(#2-#1)*0.5},0)
  node[myRec,fill=#4, minimum width=#2cm-#1cm,label=above:{\small{#3}}] {}%
}

\newcommand{\writeAbove}[2]{%
  -- ({#1},0)
  node %
  [label={[anchor=mid]above:\small{#2}}] {}%
}

\newcommand{\makeoplabel}[1]{#1\hspace{0.2em}}
\newcommand{\writesyntax}[2]{$#1$.write(#2)}
\newcommand{\readsyntax}[2]{$#1$.read$\rightarrow$#2}

\newcommand{\writeOpColor}[6]{\operation{#3}{#4}{\makeoplabel{#5}\writesyntax{#1}{#2}}{#6}}
\newcommand{\readOpColor}[6]{\operation{#3}{#4}{\makeoplabel{#5}\readsyntax{#1}{#2}\phantom{)}}{#6}}
\newcommand{\writeOp}[5]{\writeOpColor{#1}{#2}{#3}{#4}{#5}{red}}
\newcommand{\readOp}[5]{ \readOpColor {#1}{#2}{#3}{#4}{#5}{green}}

\newcommand{\processid}{}

\newcommand{\ProcessWithId}[4]{
  \renewcommand{\processid}{#1}
  \pgftransformyshift{-0.9cm}
  \draw (0,0) node[rectangle,fill=white,anchor=east] (#1) {#2} #4 edge[->] (#3,0) ;
}
\newcommand{\Process}[3]{%
  \ProcessWithId{#1}{#1}{#2}{#3}
}
\newcommand{\ProcessNoId}[3]{%
  \ProcessWithId{}{#1}{#2}{#3}
}

\newcommand{\procmessage}[6]{%
\draw (#1 -| #2,0) edge [-latex,#6] node [auto] {#5} (#3 -| #4,0);%
}

\DeclareMathOperator*{\argmin}{arg\,min}

\usepackage{ifthen}
\usepackage{calc}
\usepackage[usenames,dvipsnames,pdftex]{color}

\newcommand{\ftanote}[2]{\marginpar{\centering #1 \fbox{#2}}}

\newcommand{\annote}[3]{{\color{#3}%
	   \ifthenelse{\equal{#1}{}}{[\scshape #2]}{[\textsc{#2:} \emph{#1}]} \normalcolor}\ftanote{\color{#3} \sc}{#2}}

\newlength {\squarewidth}

\newcommand{\toto}{xxx}
\newenvironment{proofT}{\noindent{\bf Proof }}
{\hspace*{\fill}$\Box_{Theorem~\ref{\toto}}$\par\vspace{3mm}}
\newenvironment{proofL}{\noindent{\bf Proof }}
{\hspace*{\fill}$\Box_{Lemma~\ref{\toto}}$\par\vspace{3mm}}

\newcommand{\Xomit}[1]{}

\begin{document}

\RRNo{2022}

\RRdate{\today}

\RRtitle{La cohérence en \oe{}il de poisson : maintenir la synchronisation des données dans un monde géo-répliqué}

\RRetitle{Fisheye Consistency: Keeping Data in Synch in a Georeplicated World}


\RRauthor{
Roy Friedman\thanks{The Technion Haifa, Israel}, Michel Raynal\thanks{Institut Universitaire de France}$^,$\thanks{Université de Rennes 1, IRISA}
François Taïani\footnotemark[2]
}
\authorhead{R. Friedman, M. Raynal \& F. Taïani}




\RRresume{\small
Au cours des trente dernières années, de nombreuses conditions de cohérence pour les données répliquées ont été proposées et mises en oeuvre. Les exemples courants de ces conditions comprennent la linéarisabilité (ou atomicité), la cohérence séquentielle, la cohérence causale, et la cohérence éventuelle. Ces conditions de cohérence sont généralement définies indépendamment des entités informatiques (noeuds) qui manipulent les données répliquées; c'est à dire qu'elles ne prennent pas en compte la façon dont les entités informatiques peuvent être liées les unes aux autres, ou géographiquement distribuées. Pour combler ce manque, ce document introduit la notion de \emph{graphe de proximité} entre les noeuds de calcul d'un système réparti. Si deux noeuds sont connectés dans ce graphe, leurs activités doivent satisfaire une condition de cohérence forte, tandis que les opérations invoquées par d'autres noeuds peuvent ne satisfaire qu'une condition plus faible. Nous proposons d'utiliser un tel graphe pour fournir une approche générique à l'hybridation de conditions de cohérence des données dans un même système. Nous illustrons cette approche sur l'exemple de la cohérence séquentielle et de la cohérence causale, et présentons un modèle dans lequel, d'une part, toutes les opérations sont causalement cohérentes, et, d'autre part, les opérations par des processus qui sont voisins dans le graphe de proximité satisfont la cohérence séquentielle. Nous proposons et prouvons un algorithme distribué basé sur ce graphe de proximité, qui combine la cohérence séquentielle et la cohérence causal (nous appelons la cohérence obtenue \emph{cohérence en oeil de poisson}). Ce faisant, le papier non seulement étend le domaine des conditions de cohérence, mais fournit une solution algorithmiquement correcte et générique directement applicable aux systèmes géo--répartis modernes.
}

\RRabstract{\small
Over the last thirty years, numerous consistency conditions
for replicated data have been proposed and implemented. Popular examples of such conditions include linearizability (or atomicity), sequential consistency, causal
consistency, and eventual consistency.
These consistency conditions are usually  defined
independently from  the computing entities (nodes)
that manipulate the replicated data; i.e., they do not take into account how computing entities might be linked to one another, or geographically distributed.
To address this lack, as a first contribution, this paper introduces
the notion of {\it proximity graph} between computing nodes.
If two nodes are connected in this graph, their
operations must satisfy a strong consistency condition,
while the operations invoked by other nodes are allowed to satisfy a weaker
condition. The second contribution is the use of such a graph
to provide a generic approach to the hybridization of data
consistency conditions into the same system.
We illustrate this approach on sequential consistency and causal
consistency, and present a model in which all data operations are causally
consistent, while operations by neighboring processes in the proximity graph are sequentially consistent.
The third  contribution of the paper is the design and the proof
of a distributed algorithm based on this proximity graph,
which combines  sequential consistency and causal consistency (the resulting
condition is called {\it fisheye consistency}). In doing so the paper not only
extends the domain of consistency  conditions, but provides a
generic provably correct solution of direct relevance to modern georeplicated systems.}

\RRkeyword{
Asynchronous message-passing system,
Broadcast abstraction, Causal consistency,   Data consistency,
Data replication, Geographical distribution,
Linearizability, Provable property, Sequential consistency.
}

\RRmotcle{Systèmes par passage de messages asynchrones, Abstractions de diffusion, Cohérence Causale, Cohérence des données, Réplication des données, Distribution géographiques, Linéarisabilité, Propriété prouvables, Cohérence Séquentielle}

\newtheorem{lemma}{Lemma}
\newtheorem{theorem}{Theorem}

\makeRR

\section{Introduction}

\paragraph{Data consistency in distributed systems}

Distributed computer systems are growing in size, be it in terms of machines, data, or geographic distribution. Insuring strong
consistency guarantees (e.g., linearizability~\cite{HW90})
in such large-scale systems has attracted a lot of attention
over the years, and remains today a highly challenging area, for reasons
of cost, failures, and scalability.
One popular strategy to address these challenges has been to propose and implement weaker guarantees (e.g., causal consistency~\cite{ANBHK95}, or eventual consistency~\cite{TTPDSH95}).


These weaker consistency models are not a desirable goal in themselves~\cite{AF96}, but rather an unavoidable compromise to obtain acceptable performance and availability~\cite{AW94,Br00,XSKWYAM14}. These works try in general to minimize the violations of strong consistency, as these create anomalies for programmers and users. They further emphasize the low probability of such violations in their real deployments~\cite{DHJKLPSVV2007}.

\paragraph{Recent related works}
For brevity, we cannot name all the many weak consistency conditions that have been proposed in the past.
We focus instead on the most recent works in this area.
One of the main hurdles in building systems and applications based on weak consistency models is how to generate an eventually consistent and meaningful image of the shared memory or storage~\cite{TTPDSH95}.
In particular, a paramount sticking point is how to handle conflicting concurrent write (or update) operations and merge their result in a way that suits the target application.
To that end, various conditions that enables custom conflict resolution and a host of corresponding data-types have been proposed and implemented~\cite{ALR13,ABCH2013,BGHS13,BGYZ2014,LFKA2011,PMSL09,SPBZ11,SS05}.

Another form of hybrid consistency conditions can be found in the seminal works on \emph{release consistency}~\cite{GLLGGH90,KCZ92} and \emph{hybrid consistency}~\cite{AF98,Fr95}, which distinguish between strong and weak operations such that strong operations enjoy stronger consistency guarantees than weak operations.
Additional mechanisms and frameworks that enable combining operations of varying consistency levels have been recently proposed in the context of large scale and geo-replicated data centers~\cite{TPKBAL13,XSKWYAM14}.


%
%
%
%


\paragraph{Motivation and problem statement}
In spite of their benefits, the above consistency conditions generally ignore the
relative ``distance'' between nodes in the underlying ``infrastructure'',
where the notions of ``distance'' and
``infrastructure'' may be logical or physical, depending on the application.
This is unfortunate as distributed systems must scale out and geo-replication is becoming more common. In a geo-replicated system, the network latency and bandwidth connecting nearby servers is usually at least an order of magnitude better than what is obtained between remote servers. This means that the cost of maintaining strong consistency among nearby nodes becomes affordable compared to the overall network costs and latencies in the system. 

Some production-grade systems acknowledge the importance of distance when enforcing consistency, and do propose consistency mechanisms based on node locations in a distributed system (e.g. whether nodes are located in the same or in different data-centers). Unfortunately these production-grade systems usually do not distinguish between semantics and implementation. Rather, their consistency model is defined in operational terms, whose full implications can be difficult to grasp. In Cassandra~\cite{LM2010}, for instance, the application can specify for each operation the type of consistency guarantee it desires. 
For example, the constraints QUORUM and ALL require the involvement of a quorum of replicas and of all replicas, respectively;
while LOCAL\_QUORUM is satisfied when a quorum of the local data center is contacted, and EACH\_QUORUM requires a quorum in each data center.
These guarantees are defined by their implementation, but do not provide the programmer with a precise image of the consistency they deliver.

The need to take into account ``distance'' into consistency models, and the current lack of any formal underpinning to do so are exactly what motivates the hybridization of consistency conditions that we propose in this paper (which we call \emph{fisheye consistency}). Fisheye consistency conditions provide strong guarantees only for operations issued at nearby servers. In particular, there are many applications where one can expect that concurrent operations on the same objects are likely to be generated by geographically nearby nodes, e.g., due to business hours in different time zones, or because these objects represent localized information, etc. In such situations, a fisheye consistency condition would in fact provide global strong consistency at the cost of maintaining only locally strong consistency.

Consider for instance a node $A$ that is ``close'' to a node $B$, but ``far'' from a node $C$,
a causally consistent read/write register will offer the same (weak)
guarantees to $A$ on the operations of $B$, as on the operations of $C$.
This may be suboptimal, as many applications could benefit from varying
levels of consistency conditioned on ``how far'' nodes are from each other.
Stated differently: a node can accept that ``remote'' changes only reach
it with weak guarantees (e.g., because information takes time to travel),
but it wants changes ``close'' to it to come with strong guarantees
(as ``local'' changes might impact it more directly).

In this work, we propose to address this problem by integrating a notion of  {\it node proximity} in the
definition of {\it data consistency}.
To that end, we  formally define a new family
of hybrid consistency models that links the strength of data consistency
with the proximity of the participating nodes. In our approach, a
particular hybrid model takes as input a proximity graph, and
two consistency conditions, taken from a set of totally ordered consistency
conditions, namely a strong one and a weaker one.
A classical set of totally ordered conditions is the following one:
linearizability, sequential consistency, causal consistency, and
PRAM-consistency~\cite{LS88}.
Moreover, as already said, the notion of  proximity can be geographical
(cluster-based physical distribution of the nodes), or purely logical
(as in some peer-to-peer systems).

The philosophy we advocate is related to that of
 Parallel Snapshot Isolation (PSI) proposed in~\cite{SPAL11}. PSI combines
strong consistency (Snapshot Isolation) for transactions started at nodes in the same site
of a geo-replicated system, but only ensures causality among transactions started at different
sites. In addition, PSI prevents write-write conflicts by preventing concurrent transactions
with conflicting write sets, with the exception of commutable objects.

Although PSI and our work operate at different granularities (fisheye-consistency is expressed on individual operations, each accessing a single object, while PSI
addresses general transactions), they both show the interest of
consistency conditions
in which nearby nodes enjoy stronger semantics than remote ones.
In spite of this similitude, however, the family of consistency conditions we propose distinguishes itself from PSI in a number of key dimensions. First, PSI is a specific condition while fisheye-consistency
offers a general framework for defining multiple such conditions.
PSI only distinguished between nodes at the same physical site and remote nodes, whereas
fisheye-consistency accepts arbitrary proximity graphs, which can be physical or logical.
Finally, the definition of PSI is given in~\cite{SPAL11} by a reference implementation, whereas fisheye-consistency is defined in functional terms 
as restrictions on the ordering of operations that can be seen by applications, independently of the implementation we propose.
As a result, we believe that our formalism makes it easier for users to express and understand the semantics of a given consistency condition
and to prove the correctness of a program written w.r.t. such a condition.

\paragraph{Roadmap}
The paper is composed of~\ref{sec:conclusion} sections.
Section~\ref{sec:model} introduces the system model
and two classical data consistency conditions, namely,
sequential consistency  (SC) \cite{L79} and causal consistency (CC) \cite{ANBHK95}.
Then, Section~\ref{sec:fisheye-definition} defines the
notion of \emph{proximity graph} and the associated fisheye consistency condition,
which considers SC as its strong condition and CC as its weak condition.
Section~\ref{sec:broadcast-abstraction} presents a broadcast abstraction,
and Section~\ref{sec:fisheye-algorithm}  builds  on top of this communication
abstraction a  distributed algorithm implementing this hybrid proximity-based
data consistency condition.
These algorithms are  generic, where the genericity parameter is the
proximity graph. Interestingly, their two extreme instantiations provide
natural implementations of SC  and CC.
Finally, Section~\ref{sec:conclusion} concludes the paper.

\section{System Model and Basic Consistency Conditions}
\label{sec:model}

\subsection{System model}
The system consists of $n$  processes denoted $p_1$, ..., $p_n$. We note $\Pi$ the set of all processes.
Each process is sequential and asynchronous. ``Asynchronous'' means that
each process proceeds at its own speed, which  is arbitrary, may vary
with time, and remains always unknown to the other processes.
Said differently, there is no notion of a global time that could be
used by the processes.

Processes communicate by sending and receiving messages through
channels. Each channel is  reliable (no message loss, duplication,
creation, or corruption), and asynchronous  (transit times are
arbitrary but finite, and remain unknown to the processes).
Each pair of processes is connected by a bi-directional channel.

\subsection{Basic notions and definitions}

This section is a short reminder of the fundamental  notions typically used to define
the consistency guarantees of distributed objects, namely, operation, history,
partial order on operations, and history equivalence. Interested readers
will find in-depth presentations of these notions in
textbooks such as~\cite{AW04,HS08,L96,R13-Conc}.

\paragraph{Concurrent objects with sequential specification}
A concurrent object is an object that can be simultaneously accessed
by different processes.  At the application level the processes interact
through concurrent  objects~\cite{HS08,R13-Conc}.
Each object is defined by a sequential specification, which is a set
including all  the correct sequences of operations and their results that can be applied to and obtained from the object. These sequences are called  {\it legal} sequences.

\paragraph{Execution history}
The execution of a set of processes interacting through objects is
captured by a \emph{history} $\widehat{H}=(H,\rightarrow_H)$,
where $\rightarrow_H$ is a partial order on the set $H$ of the object
operations invoked by the processes.


\paragraph{Concurrency and sequential history}
If two operations are not ordered in a history, they are said to be
\emph{concurrent}.
A history is said to be {\it sequential} if it does not include any concurrent operations.
In this case, the partial order  $\rightarrow_H$ is a total order.

\paragraph{Equivalent history}
Let  $\widehat{H}|p$ represent the projection of $\widehat{H}$ onto
the process $p$, i.e., the restriction of $\widehat{H}$ to operations
 occurring at process $p$. Two histories $\widehat{H}_1$ and $\widehat{H}_2$
are  {\it equivalent} if no process can distinguish them, i.e.,
$\forall p \in \Pi : \widehat{H}_1|p = \widehat{H}_2|p.$


\paragraph{Legal history}
 $\widehat{H}$ being a sequential history, let
 $\widehat{H}|X$ represent  the projection of $\widehat{H}$ onto
the object $X$. A history  $\widehat{H}$ is {\it legal} if, for any object
$X$, the sequence  $\widehat{H}|X$ belongs to the specification of $X$.

\paragraph{Process Order}
Notice that since we assumed that processes are sequential, we restrict the
discussion in this paper to execution histories $\widehat{H}$ for which for every
process $p$, $\widehat{H}|p$ is sequential.
This total order is also called the \emph{process order} for $p$.

\subsection{Sequential consistency}
Intuitively, an execution is sequentially consistent if it could have been
produced by executing (with the help of a scheduler) the processes on
a monoprocessor. Formally,
a history  $\widehat{H}$ is {\it sequentially consistent} (SC) if
there exists a history $\widehat{S}$ such that:
\begin{itemize}
\item $\widehat{S}$ is sequential,
\item $\widehat{S}$ is legal (the specification of each object is respected),
\item $\widehat{H}$ and $\widehat{S}$  are equivalent
(no process can distinguish $\widehat{H}$---what occurred---and
$\widehat{S}$---what we would like to see, to be able to reason about).
\end{itemize}
One can notice that SC does not demand that
the sequence $\widehat{S}$   respects the
real-time occurrence order on the operations.
This is the fundamental difference between linearizability and SC.

\begin{figure}[tbh]
\centering
\begin{tikzpicture}
  \Process{$p$}{7}{
    \readOp {X}{0}{2.5}{3.5}{\hspace{-2em}op$_{p}^1$:}
    \writeOp{X}{3}{4.5}{6.5}{op$_{p}^2$:}
  }
  \Process{$q$}{7}{
    \writeOp {X}{2}{1.25}{2.25}{\hspace{1em}op$_{q}^1$:}
    \readOp  {X}{3}{4}{6}{op$_{q}^2$:}
  }
\end{tikzpicture}
\caption{A sequentially consistent execution}
\label{fig:example:seqCons}
\end{figure}
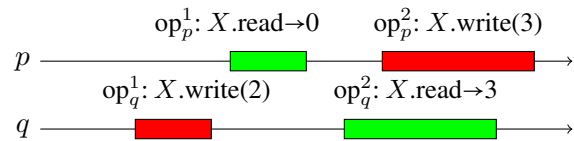

An example of a history $\widehat{H}$ that is sequentially consistent
is shown in Figure~\ref{fig:example:seqCons}.  Let us observe that,
although op$_{q}^1$ occurs before  op$_{p}^1$ in physical time,
 op$_{p}^1$ does not see the effect of the write operation op$_{q}^1$,
and still returns $0$. A legal sequential history  $\widehat{S}$,
equivalent to $\widehat{H}$, can be easily built, namely,
$X.\mbox{read}\rightarrow 0, ~ X.\mbox{write}(2), ~ X.\mbox{write}(3),
~X.\mbox{read}\rightarrow 3.$

\subsection{Causal  consistency}
\label{sec:causal-consistency}

In a sequentially consistent execution, all processes perceive all operations in the same order, which is captured by the existence of a sequential and legal history $\widehat{S}$. Causal consistency~\cite{ANBHK95} relaxes this constraint for read-write registers, and allows different processes to perceive different orders of operations, as long as causality is preserved.

Formally, a history $\widehat{H}$ in which processes interact through concurrent read/write registers is causally consistent (CC) if:
\begin{itemize}
\item There is a causal order $\leadsto_{H}$ on the operations of
$\widehat{H}$, i.e., a partial order that links each read to at most
one latest write (or otherwise to an initial value $\bot$), so that
the value returned by the read is the one written by this latest write
and $\leadsto_{H}$ respects the process order of all processes.
\item For each process $p_i$, there is a sequential and legal
history $\widehat{S}_i$ that
\begin{itemize}
\item
is equivalent to $\widehat{H}{|(p_i+W)}$,
where $\widehat{H}{|(p_i+W)}$ is the sub-history of $\widehat{H}$ that
contains all operations of $p_i$, plus the  writes of all the other processes,
\item
respects $\leadsto_{H}$ (i.e., $\leadsto_{H}\;\subseteq\; \rightarrow_{S_i}$).
\end{itemize}
\end{itemize}
Intuitively, this definition  means that all processes see causally
related write operations in the same order, but can see operations
that are not causally related
($o_1 \not\leadsto_{H} o_2 \wedge o_2 \not\leadsto_{H} o_1$)
in different orders.

\begin{figure}[tbh]
\centering
\begin{tikzpicture}
  \Process{$p$}{7}{ \writeOp{X}{2}{1   }{2   }{op$_{p}^1$:} }
  \Process{$q$}{7}{ \writeOp{X}{3}{3   }{4   }{op$_{q}^1$:} }
  \Process{$r$}{7}{ \readOp {X}{2}{1.5 }{2.5 }{op$_{r}^1$:}
                    \readOp {X}{3}{4.5 }{5.5 }{op$_{r}^2$:} }
  \Process{$s$}{7}{ \readOp {X}{3}{2.5 }{3.5 }{op$_{s}^1$:}
                    \readOp {X}{2}{5.25}{6.25}{op$_{s}^2$:} }
\end{tikzpicture}
\caption{An execution  that is causally consistent
(but not sequentially consistent)}
\label{fig:example:CausCons}
\end{figure}
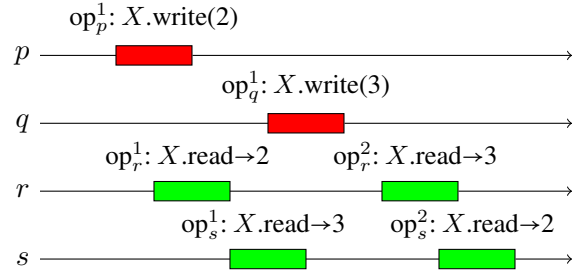

An example of causally consistent execution is given in
Figure~\ref{fig:example:CausCons}.
The processes $r$ and $s$ observe the write operations on $X$ by $p$ (op$_{p}^1$) and $q$ (op$_{q}^1$)
in two different orders. This is acceptable in a causally consistent history
because op$_{p}^1$ and op$_{q}^1$ are not causally related. This would
not be acceptable in a sequentially consistent history, where the same
total order on operations must be observed by all the  processes.
(When considering read/write objects, this constitutes the maim difference
between SC and CC.)

\section{The Family of Fisheye Consistency Conditions}
\label{sec:fisheye-definition}
This section introduces a hybrid consistency model based on
(a) two consistency  conditions and  (b) the notion of a proximity graph
defined on the  computing  nodes (processes).
The two consistency conditions must be totally ordered in the sense
that any execution satisfying the stronger one also satisfies the weaker one.
Linearizability and SC define such a pair of consistency conditions, and similarly
SC and CC are such a pair.

\subsection{The notion of a proximity graph}
Let us assume that for physical or logical reasons linked to the application,
each  process (node) can be considered either close to or remote from other processes.
This  notion of ``closeness'' can be captured trough a
 {\it proximity graph} denoted ${\cal G} = (\Pi, E_{\cal G} \subseteq \Pi\times\Pi)$, whose vertices are the $n$
processes of the system ($\Pi$). The edges are  undirected. $N_{\mathcal{G}}(p_i)$ denotes the neighbors of $p_i$ in $\mathcal{G}$.

The aim of ${\cal G}$ is to state the level of consistency imposed
on processes in the following sense: the existence of an edge between two processes in  ${\cal G}$
imposes a stronger data consistency level than between processes not
connected in ${\cal G}$.





\newcommand{\paris}{\mathit{paris}}
\newcommand{\berlin}{\mathit{berlin}}
\newcommand{\newYork}{\mathit{new}\text{-}\mathit{york}}

\paragraph{Example}

To illustrate the semantic of $\mathcal{G}$, we extend the original scenario that Ahamad, Niger \emph{et al} use to motivate causal consistency in \cite{ANBHK95}. Consider the three processes of Figure~\ref{fig:example:SC:cons:no:need:sync}, $\paris$, $\berlin$, and $\newYork$. Processes~$\paris$ and $\berlin$ interact closely with one another and behave symmetrically : they concurrently write the shared variable $X$, then set the flags $R$ and $S$ respectively to $1$, and finally read $X$. By contrast, process $\newYork$ behaves sequentially w.r.t. $\paris$ and $\berlin$: $\newYork$ waits for $\paris$ and $\berlin$ to write on $X$, using the flags $R$ and $S$, and then writes $X$.

\begin{figure}[tbh]
\begin{algorithmic}
\parbox{0.25\linewidth}{
  \ProcessIs{$\paris$}
    \State $X \leftarrow 1$
    \State $R \leftarrow 1$
    \State $a \leftarrow X$
  \EndProcessIs
}
\parbox{0.25\linewidth}{
  \ProcessIs{$\berlin$}
    \State $X \leftarrow 2$
    \State $S \leftarrow 1$
    \State $b \leftarrow X$
  \EndProcessIs
}
\parbox{0.4\linewidth}{
  \ProcessIs{$\newYork$}
    \RepeatUntilOneLine{$c \leftarrow R$}{$c=1$}
    \RepeatUntilOneLine{$d \leftarrow S$}{$d=1$}
    \State $X \leftarrow 3$
  \EndProcessIs
}
\end{algorithmic}
\caption{$\newYork$ does not need to be closely synchronized with $\paris$ and $\berlin$, calling for a hybrid form of consistency}
\label{fig:example:SC:cons:no:need:sync}
\end{figure}

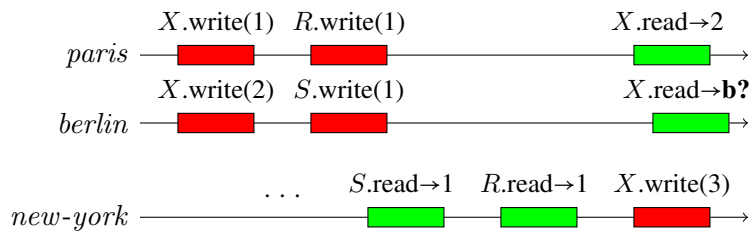
\begin{figure}[tbh]
\centering
\begin{tikzpicture}
  \renewcommand{\makeoplabel}[1]{} 


  \ProcessNoId{$\paris$}{8}{
    \writeOp{X}{1}{0.5 }{1.5 }{op$^{1}_{p}$:}
    \writeOp{R}{1}{2.25}{3.25}{op$^{2}_{p}$:}
    \readOp {X}{2}{6.5 }{7.5 }{op$^{3}_{p}$:}
  }
  \ProcessNoId{$\berlin$}{8}{
    \writeOp{X}{2}{0.5 }{1.5 }{op$^1_{q}$:}
    \writeOp{S}{1}{2.25}{3.25}{op$^2_{q}$:}
    \readOp {X}{\textbf{b?}}{6.75}{7.75}{op$^{3}_{q}$:}
  }
  \pgftransformyshift{-0.35cm}

  \ProcessNoId{$\newYork$}{8}{
    \writeAbove{1.875}{. . .}
    \readOp {S}{1}{3   }{4   }{op$^1_{t}$:}
    \readOp {R}{1}{4.75}{5.75}{op$^2_{t}$:}
    \writeOp{X}{3}{6.5 }{7.5 }{op$^3_{t}$:}
  }
\end{tikzpicture}
\caption{Executing the program of Figure~\ref{fig:example:SC:cons:no:need:sync}.}
\label{fig:example:SC:usefulness}
\end{figure}

If we assume a model that provides causal consistency at a minimum, the write of $X$ by $\newYork$ is guaranteed to be seen after the writes of $\paris$ and $\berlin$ by all processes (because $\newYork$ waits on $R$ and $S$ to be set to $1$). Causal consistency however does not impose any consistent order on the writes of $\paris$ and $\berlin$ on $X$. In the execution shown on Figure~\ref{fig:example:SC:usefulness}, this means that although $\paris$ reads $2$ in $X$ (and thus sees the write of $\berlin$ after its own write), $\berlin$ might still read $1$ in $b$ (thus perceiving `$X$.write(1)' and `$X$.write(2)' in the opposite order to that of $\paris$).

Sequential consistency removes this ambiguity: in this case, in Figure~\ref{fig:example:SC:usefulness}, $\berlin$ can only read $2$ (the value it wrote) or $3$ (written by $\newYork$), but not $1$. Sequential consistency is however too strong here: because the write operation of $\newYork$ is already causally ordered with those of $\paris$ and $\berlin$, this operation does not need any additional synchronization effort.
This situation can be seen as an extension of the \emph{write concurrency freedom} condition introduced in \cite{ANBHK95}: $\newYork$ is here free of concurrent write w.r.t. $\paris$ and $\berlin$, making causal consistency equivalent to sequential consistency for $\newYork$. $\paris$ and $\berlin$ however write to $X$ concurrently, in which case causal consistency is not enough to ensure strongly consistent results.

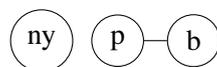
\begin{figure}[tb]
\centering
\tikzstyle{process}=[circle,minimum size=6.5mm,draw]
\begin{tikzpicture}
  \node[process] (p)  {p};
  \node[process] (b)  [right of=p] {b};
  \node[process] (ny) [left of=p] {ny};
  \draw (p) -- (b) ;
\end{tikzpicture}
\caption{Capturing the synchronization needs of Fig.~\ref{fig:example:SC:cons:no:need:sync} with a proximity graph $\mathcal{G}$}
\label{fig:proxi:graph:for:example}
\end{figure}

If we assume $\paris$ and $\berlin$ execute in the same data center, while $\newYork$ is located on a distant site, this example illustrates  a more general case in which, because of a program's logic or activity patterns, no operations at one site ever conflict with those at another. In such a situation, rather than enforce a strong (and costly) consistency in the whole system, we propose a form of consistency that is strong for processes within the same site (here $\paris$ and $\berlin$), but weak between sites (here between ${\paris,\berlin}$ on one hand and $\newYork$ on the other).

In our model, the synchronization needs of individual processes are captured by the \emph{proximity graph} $\mathcal{G}$ introduced at the start of this section and shown in Figure~\ref{fig:proxi:graph:for:example}: $\paris$ and $\berlin$ are connected, meaning the operations they execute should be perceived as strongly consistent w.r.t. one another ; $\newYork$ is neither connected to $\paris$ nor $\berlin$, meaning a weaker consistency is allowed between the operations executed at $\newYork$ and those of $\paris$ and $\berlin$.




\subsection{Fisheye consistency for the pair
(sequential consistency, causal consistency)}
\label{sec:fish-cons-pair}

When applied to the scenario of Figure~\ref{fig:example:SC:usefulness}, fisheye consistency  combines two consistency conditions (a strong and a weaker one, here causal and sequential consistency) and a proximity graph to form an hybrid distance-based consistency condition, which we call \emph{$\mathcal{G}$-fisheye (SC,CC)-consistency}.


The intuition in combining SC and CC  is to require that (write) operations
be observed in the same order by all processes if:
\begin{itemize}
\item They are causally related (as in causal consistency),
\item Or they occur on ``close'' nodes (as defined  by  $\mathcal{G}$).
\end{itemize}

\newcommand{\leadstostar}{\overset{\filledstar}{\leadsto}}

\paragraph{Formal definition}
Formally, we say that a history $\widehat{H}$ is $\mathcal{G}$-fisheye
 (SC,CC)-consistent if:
\begin{itemize}
\item There is a causal order $\leadsto_{H}$ induced by $\widehat{H}$
(as in causal consistency); and
\item $\leadsto_{H}$ can be extended to a 
 subsuming order
$\leadstostar_{H,\mathcal{G}}$
(i.e. $\leadsto_{H}\;\subseteq\;
                  \leadstostar_{H,\mathcal{G}}$)
so that
$$\forall p,q \in \mathcal{G} :\:
 (\leadstostar_{H,\mathcal{G}}){|\{p,q\}}
  \textrm{ is a total order}$$
where $(\leadstostar_{H,\mathcal{G}}){|(\{p,q\}\cap W)}$
is the restriction of
$\leadstostar_{H,\mathcal{G}}$ to the write operations
of $p$ and $q$; and
\item for each process $p_i$  there is a history $\widehat{S}_i$ that
\begin{itemize}
\item    (a) is sequential and legal;
\item    (b) is equivalent to $\widehat{H}{|(p_i+W)}$; and
\item    (c) respects $\leadstostar_{H,\mathcal{G}}$,
 i.e.,
 $(\leadstostar_{H,\mathcal{G}})|(p_i+W) \subseteq\;
 (\rightarrow_{S_i})$.
\end{itemize}
\end{itemize}



If we apply this definition to the example of Figure~\ref{fig:example:SC:usefulness} with the proximity graph proposed in Figure~\ref{fig:proxi:graph:for:example} we obtain the following: because $\paris$ and $\berlin$ are connected in $\mathcal{G}$, $X.\text{write}(1)$ by $\paris$ and $X.\text{write}(2)$ by $\berlin$ must be totally ordered in $\leadstostar_{H,\mathcal{G}}$ (and hence in any sequential history $\widehat{S}_i$ perceived by any process $p_i$). $X.\text{write}(3)$ by $\newYork$ must be ordered after the writes on $X$ by $\paris$ and $\berlin$ because of the causality imposed by $\leadsto_H$. As a result,  if the system is $\mathcal{G}$-fisheye
(SC,CC)-consistent, \textbf{b?} can be equal to $2$ or $3$, but not to $1$. This set of possible values is as in sequential consistency, with the difference that $\mathcal{G}$-fisheye
(SC,CC)-consistency does not impose any total order on the operation of $\newYork$.

Given a system of $n$ processes,
let $\varnothing$ denote the graph ${\cal G}$ with no edges, and
 $K$ denote the graph  ${\cal G}$ with an edge connecting each pair of
distinct processes.  It is easy to see that CC is
$\varnothing$-fisheye (SC,CC)-consistency. Similarly SC is $K$-fisheye (SC,CC)-consistency.





\begin{figure}[tb]
\centering
\begin{tikzpicture}[node distance=0.9cm]

  \Process{$p$}{10}{  \writeOp     {X}{2}{0.5 }{1.5 }{op$_{p}^1$:}
                      \writeOpColor{Y}{4}{5   }{6   }{op$_{p}^2$:}{orange} }

  \Process{$q$}{10}{  \writeOp     {X}{3}{2   }{3   }{op$_{q}^1$:}
                      \readOpColor {Y}{4}{6   }{7   }{op$_{q}^2$:}{blue}
                      \readOpColor {Y}{5}{8.5 }{9.5 }{op$_{q}^3$:}{blue} }

  \Process{$r$}{10}{  \readOp      {X}{2}{1   }{2   }{op$_{r}^1$:}
                      \readOp      {X}{3}{3.75}{4.75}{op$_{r}^2$:}
                      \writeOpColor{Y}{5}{6.5 }{7.5 }{op$_{r}^3$:}{orange} }

  \Process{$s$}{10}{  \readOp      {X}{3}{1.5 }{2.5 }{\hspace{-2em}op$_{s}^1$:}
                      \readOp      {X}{\textbf{x?}}{3.75}{ 4.75}{op$_{s}^2$:}
                      \readOpColor {Y}{5}{6.25}{ 7.25}{op$_{s}^3$:}{blue}
                      \readOpColor {Y}{\textbf{y?}}{8.75}{ 9.75}{op$_{s}^4$:}{blue} }


  \node[circle,draw] (q) [left of=$q$] {q};
  \node[circle,draw] (p) [above of=q] {p};
  \node[circle,draw] (r) [below of=q] {r};
  \node[circle,draw] (s) [below of=r] {s};
  \draw (p) -- (q) (r) -- (s);

\end{tikzpicture}
\caption{Illustrating $\mathcal{G}$-fisheye (SC,CC)-consistency}
\label{fig:example:HybridSeqConsCausCons}
\end{figure}
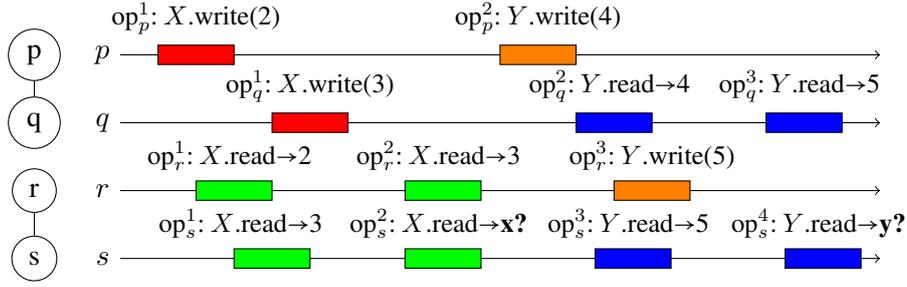

\paragraph{A larger example}
Figure~\ref{fig:example:HybridSeqConsCausCons} and
Table~\ref{tab:possible-exec-fisheye3} illustrate the semantic of
$\mathcal{G}$-fisheye (SC,CC) consistency on a second, larger, example.
In this example, the processes $p$ and $q$ on one hand,
and $r$ and $s$ on the other hand, are neighbors in the proximity graph $\mathcal{G}$ (shown on the left).
There are two pairs of write operations: $\textrm{op}^1_p$
and $\textrm{op}^1_q$ on the register $X$, and $\textrm{op}^2_p$
and $\textrm{op}^3_r$ on the register $Y$.
In a sequentially consistency history, both pairs of writes must be
seen in the same order by all processes.
As a consequence, if $r$ sees the value $2$ first ($\textrm{op}^1_r$)
and then the value $3$ ($\textrm{op}^2_r$) for $X$, $s$ must do the same,
and only the value $3$ can be returned by \textbf{x?}.
For the same reason, only the value $3$ can be returned by \textbf{y?}, as shown
in the first line of Table~\ref{tab:possible-exec-fisheye3}.

\begin{table}[tb]
\caption{Possible executions for the history of
Figure~\ref{fig:example:HybridSeqConsCausCons}}
\label{tab:possible-exec-fisheye3}
\begin{center}
\begin{tabular}{lcc}
 \textbf{Consistency}                   &  \textbf{x?}  &  \textbf{y?}  \\
\hline
 Sequential Consistency                 &  3            &  5         \\
 Causal Consistency                     &  \{2,3\}      &  \{4,5\}   \\
 $\mathcal{G}$-fisheye (SC,CC)-consistency  &  3            &  \{4,5\}   \\
\end{tabular}
\end{center}
\end{table}

In a causally consistent history, however, both pairs of writes
($\{\textrm{op}^1_p,\textrm{op}^1_q\}$ and
$\{\textrm{op}^2_p,\textrm{op}^3_r\}$) are causally independent.
As a result, any two processes can see each pair in
different orders. \textbf{x?} may return 2 or 3, and
\textbf{y?} 4 or 5 (second line of Table~\ref{tab:possible-exec-fisheye3}).

$\mathcal{G}$-fisheye (SC,CC)-consistency provides intermediate guarantees:
 because $p$ and $q$ are neighbors in $\mathcal{G}$,
$\textrm{op}^1_p$ and $\textrm{op}^1_q$ must be observed in the same order
 by all processes. \textbf{x?} must return 3, as in a sequentially
consistent history. However, because $p$ and $r$ are not connected in
$\mathcal{G}$, $\textrm{op}^2_p$ and $\textrm{op}^3_r$ may be seen in
different orders by different processes (as in a causally consistent history),
and \textbf{y?} may return 4 or 5 (last line of
Table~\ref{tab:possible-exec-fisheye3}).

\section{Construction of an Underlying (SC,CC)-Broadcast Operation}
\label{sec:broadcast-abstraction}

Our implementation of $\mathcal{G}$-fisheye (SC,CC)-consistency relies on a broadcast operation with hybrid ordering guarantees. In this section, we present this hybrid broadcast abstraction, before moving on the actual implementation of of $\mathcal{G}$-fisheye (SC,CC)-consistency in Section~\ref{sec:fisheye-algorithm}.


\subsection{$\mathcal{G}$-fisheye (SC,CC)-broadcast: definition}
\label{sec:mathcalg-fisheye-sc}
The hybrid broadcast we proposed, denoted $\mathcal{G}$-(SC,CC)-broadcast,
is parametrized by a proximity graph ${\cal G}$ which determines which
kind of delivery order should be applied to which messages, according
to the position of the sender in the graph ${\cal G}$.
Messages  (SC,CC)-broadcast by processes which are neighbors in
${\cal G}$ must be delivered in the same order at all the processes,
while the delivery of the other messages only need to respect causal order.

The (SC,CC)-broadcast abstraction  provides the processes with two operations,
denoted $\func{TOCO\_broadcast}()$ and $\func{TOCO\_deliver}()$.
We say that messages are  toco-broadcast and toco-delivered.

\paragraph{Causal message order}
Let $M$ be the set of messages that are toco-broadcast.
The causal message delivery order,  denoted  $\leadsto_M$,
is defined as follows~\cite{BJ87,RST91}. Let $m_1,m_2 \in M$;
$m_1 \leadsto_M  m_2$, iff one of the following conditions holds:
\begin{itemize}
\item $m_1$ and  $m_2$ have been toco-broadcast by the same process,
with $m_1$ first;
\item $m_1$ was toco-delivered by a process $p_i$ before this process
toco-broadcast $m_2$;
\item There exists a  message $m$ such that
$(m_1 \leadsto_M  m)~\wedge~(m \leadsto_M  m_2)$.
\end{itemize}

\paragraph{Definition of the $\mathcal{G}$-fisheye (SC,CC)-broadcast}
The (SC,CC)-broadcast abstraction is defined by the following properties.

\begin{description}
\item[Validity.]
If a process toco-delivers a message $m$, this message
was toco-broadcast by some process. (No spurious message.)
\item[Integrity.]
A message is toco-delivered at most once. (No duplication.)
\item[$\mathcal{G}$-delivery order.]
For all the processes $p$ and $q$ such that $(p,q)$ is an edge of ${\cal G}$,
and for all the messages $m_p$ and $m_q$ such that
 $m_p$ was toco-broadcast by $p$ and  $m_q$ was toco-broadcast by $q$,
if a process toco-delivers  $m_p$ before $m_q$, no process
toco-delivers  $m_q$ before $m_p$.
\item[Causal order.]
If $m_1 \leadsto_M  m_2$, no process toco-delivers $m_2$ before $m_1$.
\item[Termination.]
If a process toco-broadcasts a message $m$,
this message is toco-delivered by all  processes.
\end{description}
It is easy to see that if  $\mathcal{G}$ has no edges, this definition
boils down to causal delivery, and if  $\mathcal{G}$ is fully connected
(clique), this definition specifies total order delivery respecting causal
order. Finally, if  $\mathcal{G}$ is fully connected and we suppress
the ``causal order'' property, the definition boils to total order delivery.

\subsection{$\mathcal{G}$-fisheye (SC,CC)-broadcast: algorithm}

\paragraph{Local variables}
To implement the $\mathcal{G}$-fisheye (SC,CC)-broadcast abstraction,
each process $p_i$ manages three local variables.
\begin{itemize}
\item $causal_i[1..n]$ is a local vector clock used to
ensure a causal delivery order of the messages; $causal_i[j]$ is the
sequence number of the next  message that $p_i$ will toco-deliver from $p_j$.
\item $total_i[1..n]$ is a vector of logical  clock values
such that  $total_i[i]$ is the local logical clock of $p_i$ (Lamport's clock),
and  $total_i[j]$ is the value of  $total_j[j]$ as known by $p_i$.
\item $pending_i$ is a set containing the messages received and not
yet toco-delivered by $p_i$.
\end{itemize}

\paragraph{Description of the algorithm}
Let us remind that for simplicity, we assume that the channels are FIFO.
Algorithm~\ref{alg:broadcast-prot} describes the behavior of a process $p_i$.
This behavior is decomposed into four parts.

\newcommand{\scausal}{s\mathunderscore caus}
\newcommand{\stotal}{s\mathunderscore tot}

\begin{algorithm}
\caption{The $\mathcal{G}$-fisheye (SC,CC)-broadcast algorithm
executed by $p_i$}
\label{alg:broadcast-prot}

\begin{algorithmic}[1]
\setlength{\arraycolsep}{2pt}

\algrenewcommand\algorithmicindent{1.2em}

\Function{TOCO\_broadcast}{$m$}\label{line:func:tocobroadcast}

  \State \label{line:toco-bcast:inc:tot}
  $total_i[i]\leftarrow total_i[i]+1$

  \ForAllDo{$p_j \in \Pi \setminus \{p_i\}$}{$\func{send}$
    \textsc{tocobc}$(m,\langle causal_i[\cdot], total_i[i], i
    \rangle)$}
  to $p_j$\label{line:sending:tocobc:msg}

  \State \label{line:inserting:sent:m:into:pending}
  $pending_i \leftarrow pending_i \cup \big\langle m,\langle
  causal_i[\cdot], total_i[i], i \rangle \big\rangle$

  \State \label{line:toco-bcast:inc:causal}
  $causal_i[i]\leftarrow causal_i[i]+1$

\EndFunction\label{line:endfunc:tocobroadcast}

\skipSomeSpace

\OnReceive{\textsc{tocobc}$(m,\langle \scausal_j^m[\cdot],
  \stotal_j^m, j \rangle)$}\label{line:tocobc-receive}

  \State \label{line:inser:mj:pendingi}
  $pending_i \leftarrow pending_i \cup \big\langle m,\langle
  \scausal_j^m[\cdot], \stotal_j^m, j \rangle \big\rangle$

  \State \label{line:toco-receive:update:total:i:j}
  $total_i[j] \leftarrow \stotal_j^m$
  \Comment{Last message from $p_j$ had timestamp
    $\stotal_j^m$}


  \If{$total_i[i] \leq \stotal_j^m$}\label{line:toco-receive:condition}

     \State \label{line:toco-receive:inc}
     $total_i[i] \leftarrow \stotal_j^m + 1$
     \Comment{Ensuring global logical clocks}

     \ForAllDo{$p_k \in \Pi \setminus \{p_i\}$}{
       \label{line:toco-receive:inform:others}
       $\func{send}$ \textsc{catch\_up}\((total_i[i], i)\) to $p_k$
     }

  \EndIf

\EndOnReceive\label{line:end:tocobc-receive}

\skipSomeSpace

\OnReceive{\textsc{catch\_up}\((last\_date_j, j)\)}\label{line:catchup-receive}

  \State \label{line:catchup-receive:inc}
  $total_i[j] \leftarrow last\_date_j$

\EndOnReceive\label{line:catchup-receive:end}

\skipSomeSpace

\Background{$T$}\label{line:background:task:start}
  \Loop
     \WaitUntilWhere{$C \neq \varnothing$}

       \State \label{line:C:filtering}
       $C \equiv \Big\{ \big\langle m, \langle \scausal_j^m[\cdot],
       \stotal_j^m, j \rangle \big\rangle \in pending_i \;\Big|\;
       \scausal_j^m[\cdot] \leq causal_i[\cdot]\Big\}$

     \EndWaitUntilWhere

     \WaitUntilWhere{$T_1 \neq \varnothing$}
       \State \label{line:T1:filtering}
       $T_1 \equiv \left\{ \big\langle m, \langle \scausal_j^m[\cdot],
       \stotal_j^m, j \rangle \big\rangle \in C\;\middle|\; \forall
       p_k \in N_\mathcal{G}(p_j) : \langle total_i[k],k \rangle >
       \langle \stotal_j^m, j \rangle \right\}$

     \EndWaitUntilWhere

     \WaitUntilWhere{$T_2 \neq \varnothing$}

       \State \label{line:T2:filtering}
       $T_2 \equiv \left\{ \big\langle m, \langle \scausal_j^m[\cdot],
       \stotal_j^m, j \rangle \big\rangle \in T_1\;\middle|
         \begin{array}{l}
           \forall p_k \in N_\mathcal{G}(p_j),\\
           \forall \big\langle m_k, \langle \scausal_k^{m_k}[\cdot],
           \stotal_k^{m_k}, k \rangle \big\rangle \\ \hfill{}\in
           pending_i : \\
           \hspace{2em}\langle \stotal_k^{m_k}, k\rangle >  \langle  \stotal_j^m, j \rangle
       \end{array} \right\}$

     \EndWaitUntilWhere

     \State \label{line:argmin}
     $\big\langle m_0, \langle \scausal_{j_0}^{m_0}[\cdot],
     \stotal_{j_0}^{m_0}, j_0 \rangle \big\rangle \leftarrow
     \displaystyle\argmin_{\langle m, \langle
       \scausal_j^m[\cdot],\stotal_j^m, j \rangle\rangle \in
       T_2}\big\{ \langle \stotal_j^m, j \rangle
     \big\}$

     \State \label{line:remove:from:pending}
     $pending_i \leftarrow pending_i \setminus \big\langle m_0,
     \langle \scausal_{j_0}^{m_0}[\cdot],\stotal_j^m, j_0 \rangle
     \big\rangle$

     \State \label{line:deliver:tocobc}
     \Call{TOCO\_deliver}{$m_0$} to application layer

     \IfThenCommand{$j_0 \neq i$}{
       $causal_i[j_0] \leftarrow causal_i[j_0]+1}$
       \label{line:inc:causal:deliver} \Comment{for $causal_i[i]$
       see line~\ref{line:toco-bcast:inc:causal}
     }

  \EndLoop
\EndBackground\label{line:background:task:end}

\skipSomeSpace

\end{algorithmic}
\end{algorithm}

The first part
(lines~\ref{line:func:tocobroadcast}-\ref{line:endfunc:tocobroadcast})
is the code of the operation $\func{TOCO\_broadcast}(m)$. Process
$p_i$ first increases its local clock $total_i[i]$ and sends the
protocol message \textsc{tocobc}$(m,\langle causal_i[\cdot], total_i[i], i \rangle)$
to each other process.  In addition to the application message $m$,
this protocol message carries the control information needed to ensure
the correct toco-delivery of $m$, namely, the local causality vector
($causal_i[1..n]$), and the value of the local clock
($total_i[i]$). Then, this protocol message is added to the set
$pending_i$ and $causal_i[i]$ is increased by $1$ (this captures the
fact that the future application messages toco-broadcast by $p_i$ will
causally depend on $m$).

The second part
(lines~\ref{line:tocobc-receive}-\ref{line:end:tocobc-receive}) is the
code executed by $p_i$ when it receives a protocol message \textsc{tocobc}$(m,$$\langle
\scausal_j^m[\cdot],$ $\stotal_j^m,$ $j \rangle)$ from $p_j$.  When this occurs $p_i$
adds first this protocol message to $pending_i$, and updates its view
of the local clock of $p_j$ ($total_i[j]$) to the sending date of the
protocol message (namely, $\stotal_j^m$).  Then, if the local clock of
$p_i$ is late ($total_i[i] \leq \stotal_j^m$), $p_i$ catches up
  (line~\ref{line:toco-receive:inc}), and informs the other processes of it (line~\ref{line:toco-receive:inform:others}).

The third part (lines~\ref{line:catchup-receive}-\ref{line:catchup-receive:end}) is the processing of a catch up message
from a process $p_j$. In this case, $p_i$ updates its view of $p_j$'s
local clock to the date carried by the catch up message. Let us notice that,
as channels are FIFO, a view $stotal_i[j]$ can only increase.

The final part (lines~\ref{line:background:task:start}-\ref{line:background:task:end}) is a background task executed by $p_i$,
where   the  application messages  are toco-delivered.
The set $C$ contains the protocol messages  that were  received,
have not yet been  toco-delivered, and are ``minimal'' with respect
to the causality relation  $\leadsto_M$. This minimality is determined
from the vector clock $\scausal_j^m[1..n]$, and the current value
of $p_i$'s vector clock ($causal_i[1..n]$).
If only causal consistency was considered, the messages in $C$
could be delivered.

Then, $p_i$ extracts from $C$ the messages that can be toco-delivered.
Those are usually called {\it stable} messages. The notion of stability
refers here to the delivery constraint imposed by the proximity graph
${\cal G}$. More precisely, a set $T_1$ is first computed, which contains the
messages of $C$  that (thanks to the FIFO channels and the catch up
messages) cannot be made unstable (with respect to the total delivery
order defined by ${\cal G}$) by messages that $p_i$ will receive
in the future. Then the set $T_2$ is computed,  which is the subset of $T_1$
such that no message received, and not yet toco-delivered, could make
incorrect -- w.r.t. ${\cal G}$ -- the toco-delivery of a message of $T_2$.

Once a non-empty set $T_2$ has been computed, $p_i$ extracts the message
$m$  whose timestamp  $\langle  \stotal_j^m[j], j \rangle$  is ``minimal''
 with respect to the  timestamp-based total order ($p_j$ is the sender of $m$).
This message is then removed from $pending_i$ and toco-delivered.
Finally, if $j\neq i$, $causal_i[j]$ is increased to take into account
this toco-delivery (all the messages $m'$ toco-broadcast by $p_i$
in the future will be such that $m \leadsto m'$, and this is encoded
in  $causal_i[j]$). If $j=i$, this causality update was done at line~\ref{line:toco-bcast:inc:causal}.

\begin{restatable}{theorem}{algoimplementscccbroadcast}
\label{theo:proof-toco-bcast}
Algorithm~\ref{alg:broadcast-prot} implements a $\mathcal{G}$-fisheye (SC,CC)-broadcast.
\end{restatable}


\newcommand{\proofTocoBcast}{


We use the usual partial order on vector clocks:
$$C_1[\cdot] \leq C_2[\cdot] \textrm{ iff } \forall p_i \in \Pi : C_1[i] \leq C_2[i] $$
with its accompanying strict partial order:
$$C_1[\cdot] < C_2[\cdot] \textrm{ iff } C_1[\cdot] \leq C_2[\cdot] \wedge C_1[\cdot] \neq C_2[\cdot]$$

We use the lexicographic order on the scalar clocks $\langle \stotal_j, j \rangle$:
$$\langle \stotal_j, j \rangle < \langle \stotal_i, i \rangle \textrm{ iff } (\stotal_j < \stotal_i) \vee (\stotal_j = \stotal_i \wedge i < j)$$


We start by three useful lemmata on $causal_i[\cdot]$ and $total_i[\cdot]$. These lemmata establish the traditional properties expected of logical and vector clocks.

\begin{lemma}\label{lemma:causal:increasing}
The following holds on the clock values taken by $causal_i[\cdot]$:
  \begin{enumerate}
  \item The successive values taken by $causal_i[\cdot]$ in Process $p_i$ are monotonically increasing.\label{causal:increasing}
  \item The sequence of $causal_i[\cdot]$ values attached to \textsc{tocobc} messages sent out by Process $p_i$ are strictly increasing.\label{causalii:tocobc:strict:increasing}
  \end{enumerate}
\end{lemma}


\begin{proofL}\renewcommand{\toto}{lemma:causal:increasing}
Proposition~\ref{causal:increasing} is derived from the fact that the two lines that modify $causal_i[\cdot]$ (lines~\ref{line:toco-bcast:inc:causal}, and~\ref{line:inc:causal:deliver})
only increase its value.
Proposition~\ref{causalii:tocobc:strict:increasing} follows from Proposition~\ref{causal:increasing} and the fact that line~\ref{line:toco-bcast:inc:causal} insures successive \textsc{tocobc} messages cannot include identical $causal_i[i]$ values.
\end{proofL}

\begin{lemma}\label{lemma:total:ordering}
The following holds on the clock values taken by $total_i[\cdot]$:
  \begin{enumerate}
    \item The successive values taken by $total_i[i]$ in Process $p_i$ are monotonically increasing.\label{totalii:increasing}
    \item The sequence of $total_i[i]$ values included in \textsc{tocobc} and \textsc{catch\_up} messages sent out by Process $p_i$ are strictly increasing.\label{totalii:tobobc:strict:increasing}
    \item The successive values taken by $total_i[\cdot]$ in Process $p_i$ are monotonically increasing.\label{total:increasing}
  \end{enumerate}
\end{lemma}

\begin{proofL}\renewcommand{\toto}{lemma:total:ordering}
Proposition~\ref{totalii:increasing} is derived from the fact that the lines that modify $total_i[i]$ (lines~\ref{line:toco-bcast:inc:tot} and~\ref{line:toco-receive:inc}) only increase its value (in the case of line~\ref{line:toco-receive:inc} because of the condition at line~\ref{line:toco-receive:condition}).
Proposition~\ref{totalii:tobobc:strict:increasing} follows from Proposition~\ref{totalii:increasing}, and the fact that lines~\ref{line:toco-bcast:inc:tot} and~\ref{line:toco-receive:inc} insures successive \textsc{tobobc} and \textsc{catch\_up} messages cannot include identical $total_i[i]$ values.
%

To prove Proposition~\ref{total:increasing}, we first show that:
\begin{equation}\forall j \neq i : \textrm{ the successive values taken by } total_i[j]\textrm{ in }p_i\textrm{ are monotonically increasing.}\label{eqn:goal:lemma:jnoit:increase}\end{equation}

For $j \neq i$, $total_i[j]$ can only be modified at lines~\ref{line:toco-receive:update:total:i:j} and~\ref{line:catchup-receive:inc}, by values included in \textsc{tobobc} and \textsc{catch\_up} messages, when these messages are received. Because the underlying channels are FIFO and reliable, Proposition~\ref{totalii:tobobc:strict:increasing} 
implies that the sequence of $last\_date_j$ and $\stotal_j^m$ values received by $p_i$ from $p_j$ is also strictly increasing, which shows equation (\ref{eqn:goal:lemma:jnoit:increase}).




From equation (\ref{eqn:goal:lemma:jnoit:increase}) and Proposition~\ref{totalii:increasing}, we conclude that the successive values taken by the vector $total_i[\cdot]$ in $p_i$ are monotonically increasing (Proposition~\ref{total:increasing}).
\end{proofL}

\begin{lemma}\label{lemma:scausal:link:to:delivery}
Consider an execution of the protocol. The following invariant holds: for $i\neq j$, if $m$ is a message sent from $p_j$ to $p_i$, then at any point of $p_i$'s execution outside of lines~\ref{line:deliver:tocobc}-\ref{line:inc:causal:deliver}, $\scausal_j^m[j] < causal_i[j]$ iff that $m$ has been toco-delivered by $p_i$.
\end{lemma}


\begin{proofL}\renewcommand{\toto}{lemma:scausal:link:to:delivery}
We first show that if $m$ has been toco-delivered by $p_i$, then $\scausal_j^m[j] < causal_i[j]$, outside of lines~\ref{line:deliver:tocobc}-\ref{line:inc:causal:deliver}. This implication follows from the condition $\scausal_j^m[\cdot] \leq causal_i[\cdot]$ at line~\ref{line:C:filtering}, and the increment at line~\ref{line:inc:causal:deliver}.

We prove the reverse implication by induction on the protocol's execution by process $p_i$. When $p_i$ is initialized $causal_i[\cdot]$ is null:
\begin{equation}causal^0_i[\cdot] = [0 \cdots 0]\label{eqn:causal:zero}\end{equation}

\noindent because the above is true of any process, with Lemma~\ref{causalii:tocobc:strict:increasing}, we also have
\begin{equation}\scausal_j^m[\cdot] \geq [0 \cdots 0]\label{eqn:scausal:greater:zero} \end{equation}
\noindent for all message $m$ that is toco-broadcast by Process $p_j$.

(\ref{eqn:causal:zero}) and (\ref{eqn:scausal:greater:zero}) imply that there are no messages sent by $p_j$ so that  $\scausal_j^m[j] < causal^0_i[j]$, and the Lemma is thus true when $p_i$ starts.

Let us now assume that the invariant holds at some point of the execution of $p_i$. The only step at which the invariant might become violated in when $causal_i[j_0]$ is modified for $j_0 \neq i$ at line~\ref{line:inc:causal:deliver}. When this increment occurs, the condition $\scausal_{j_0}^m[j_0] < causal_i[j_0]$ of the lemma potentially becomes true for additional messages. We want to show that there is only one single additional message, and that this message is $m_0$, the message that has just been delivered at line~\ref{line:deliver:tocobc}, thus completing the induction, and proving the lemma.

For clarity's sake, let us denote $causal_i^{\circ}[j_0]$ the value of $causal_i[j_0]$ just before line~\ref{line:inc:causal:deliver}, and $causal_i^{\bullet}[j_0]$ the value just after. We have $causal_i^{\bullet}[j_0] = causal_i^{\circ}[j_0] + 1$.

We show that $\scausal_{j_0}^{m_0}[j_o] = causal_i^{\circ}[j_0]$, where $\scausal_{j_0}^{m_0}[\cdot]$ is the causal timestamp of the message $m_0$ delivered at line~\ref{line:deliver:tocobc}. Because $m_0$ is selected at line~\ref{line:argmin}, this implies that $m_0 \in T_2 \subseteq T_1 \subseteq C$. Because $m_0 \in C$, we have
\begin{equation}\scausal_{j_0}^{m_0}[\cdot] \leq causal_i^{\circ}[\cdot]\end{equation}
\noindent at line~\ref{line:C:filtering}, and hence
\begin{equation}\scausal_{j_0}^{m_0}[j_0] \leq causal_i^{\circ}[j_0]\label{eqn:scausal:less}\end{equation}

\noindent At line~\ref{line:C:filtering}, $m_0$ has not been yet delivered (otherwise it would not be in $pending_i$). Using the contrapositive of our induction hypothesis, we have
\begin{equation}\scausal_{j_0}^{m_0}[j_0] \geq causal_i^{\circ}[j_0]\label{eqn:scausal:more}\end{equation}

\noindent (\ref{eqn:scausal:less}) and (\ref{eqn:scausal:more}) yield
\begin{equation}\scausal_{j_0}^{m_0}[j_0] = causal_i^{\circ}[j_0]\label{eqn:scausal:equal}\end{equation}

Because of line~\ref{line:toco-bcast:inc:causal}, $m_0$ is the only message $\func{tobo\_broadcast}$ by $P_{j_0}$ whose causal timestamp verifies (\ref{eqn:scausal:equal}). From this unicity and (\ref{eqn:scausal:equal}), we conclude that after $causal_i[j_0]$ has been incremented at line~\ref{line:inc:causal:deliver}, if a message $m$ sent by $P_{j_0}$ verifies $\scausal_{j_0}^m[j_0] < causal_i^{\bullet}[j_0]$, then
\begin{itemize}
\item either $\scausal_{j_0}^m[j_0] < causal_i^{\bullet}[j_0] - 1 = causal_i^{\circ}[j_0]$, and by induction assumption, $m$ has already been delivered;
\item or $\scausal_{j_0}^m[j_0] = causal_i^{\bullet}[j_0] - 1 < causal_i^{\circ}[j_0]$, and $m = m_0$, and $m$ has just been delivered at line~\ref{line:deliver:tocobc}.
\end{itemize}
\end{proofL}

\subsubsection*{Termination}

\begin{theorem}\label{theo:termination}
  All messages toco-broadcast using Algorithm~\ref{alg:broadcast-prot} are eventually toco-delivered by all processes in the system.
\end{theorem}

\begin{proofT}\renewcommand{\toto}{theo:termination}
We show Termination by contradiction. Assume a process $p_i$ toco-broadcasts a message $m_i$ with timestamp $\langle \scausal_i^{m_i}[\cdot], \stotal_i^{m_i}, i \rangle$, and that $m_i$ is never toco-delivered by $p_j$.

If $i\neq j$, because the underlying communication channels are reliable, $p_j$ receives at some point the \textsc{tocobc} message containing $m_i$ (line~\ref{line:tocobc-receive}), after which we have

\begin{equation}\big\langle m_i,\langle \scausal_i^{m_i}[\cdot], \stotal_i^{m_i}, i \rangle \big\rangle \in pending_j\label{eqn:termination:mi:in:pending}\end{equation}

If $i=j$, $m_i$ is inserted into $pending_i$ immediately after being toco-broadcast (line~\ref{line:inserting:sent:m:into:pending}), and (\ref{eqn:termination:mi:in:pending}) also holds.

$m_i$ might never be toco-delivered by $p_j$ because it never meets the condition to be selected into the set $C$ of $p_j$ (noted $C_j$ below) at line~\ref{line:C:filtering}. We show by contradiction that this is not the case. First, and without loss of generality, we can choose $m_i$ so that it has a minimal causal timestamp $\scausal_i^{m_i}[\cdot]$ among all the messages that $j$ never toco-delivers (be it from $p_i$ or from any other process). Minimality means here that
\begin{equation}\forall m_x, p_j\textrm{ never delivers }m_x \Rightarrow \neg (\scausal_x^{m_x} < \scausal_i^{m_i})\label{eqn:minimality:mi}\end{equation}

\noindent Let us now assume $m_i$ is never selected into $C_j$, i.e., we always have

\begin{equation}\neg (\scausal_i^{m_i}[\cdot] \leq causal_j[\cdot])\end{equation}

\noindent This means there is a process $p_k$ so that

\begin{equation}\scausal_i^{m_i}[k] > causal_j[k]\label{eqn:termination:k:gt:in:mi}\end{equation}

If $i=k$, we can consider the message $m'_i$ sent by i just before $m_i$ (which exists since the above implies $\scausal_i^{m_i}[i] > 0$). We have $\scausal_i^{m'_i}[i] = \scausal_i^{m_i}[i]-1$, and hence from (\ref{eqn:termination:k:gt:in:mi}) we have

\begin{equation}\scausal_i^{m'_i}[i] \geq causal_j[k]\label{eqn:termination:k:ge:in:miprime}\end{equation}

\noindent Applying Lemma~\ref{lemma:scausal:link:to:delivery} to (\ref{eqn:termination:k:ge:in:miprime}) implies that $p_j$ never toco-delivers $m'_i$ either, with $\scausal_i^{m'_i}[i] < \scausal_i^{m_i}[i]$ (by way of Proposition~\ref{causalii:tocobc:strict:increasing} of Lemma~\ref{lemma:causal:increasing}), which contradicts (\ref{eqn:minimality:mi}).

If $i\neq k$, applying Lemma~\ref{lemma:scausal:link:to:delivery} to $causal_i[\cdot]$ when $p_i$ toco-broadcasts $m_i$ at line~\ref{line:sending:tocobc:msg}, we find a message $m_k$ sent by $p_k$ with $\scausal_k^{m_k}[k] = \scausal_i^{m_i}[k]-1$ such that $m_k$ was received by $p_i$ before $p_i$ toco-broadcast $m_i$. In other words, $m_k$ belongs to the causal past of $m_i$, and because of the condition on $C$ (line~\ref{line:C:filtering}) and the increment at line~\ref{line:inc:causal:deliver}, we have

\begin{equation}\scausal_k^{m_k}[\cdot] < \scausal_i^{m_i}[\cdot]\label{eqn:term:mk:smaller}\end{equation}

\noindent As for the case $i=k$, (\ref{eqn:termination:k:gt:in:mi}) also implies

\begin{equation}\scausal_k^{m_k}[k] \geq causal_j[k]\label{eqn:termination:k:ge:in:mk}\end{equation}

\noindent which with Lemma~\ref{lemma:scausal:link:to:delivery} implies that that $p_j$ never delivers the message $m_k$ from $p_k$, and with (\ref{eqn:term:mk:smaller}) contradicts $m_i$'s minimality (\ref{eqn:minimality:mi}).

We conclude that if a message $m_i$ from $p_i$ is never toco-delivered by $p_j$, after some point $m_i$ remains indefinitely in $C_j$

\begin{equation}m_i \in C_j\end{equation}

Without loss of generality, we can now choose $m_i$ with the smallest total order timestamp $\langle \stotal_i^{m_i}, i \rangle$ among all the messages never delivered by $p_j$. Since these timestamps are totally ordered, and no timestamp is allocated twice, there is only one unique such message.

We first note that because channels are reliable, all processes $p_k \in N_\mathcal{G}(p_i)$ eventually receive the \textsc{tocobc} protocol message of $p_i$ that contains $m_i$ (line~\ref{line:tocobc-receive} and following). Lines~\ref{line:toco-receive:condition}-\ref{line:toco-receive:inc} together with the monotonicity of $total_k[k]$ (Proposition~\ref{totalii:increasing} of Lemma~\ref{lemma:total:ordering}), insure that at some point all processes $p_k$ have a timestamp $total_k[k]$ strictly larger than $\stotal_i^{m_i}$:

\begin{equation}\forall p_k \in N_\mathcal{G}(p_i): total_k[k] > \stotal_i^{m_i}\end{equation}

Since all changes to $total_k[k]$  are systematically rebroadcast to the rest of the system using \textsc{tocobc} or \textsc{catchup} protocol messages (lines~\ref{line:toco-bcast:inc:tot} and \ref{line:toco-receive:inc}), $p_j$ will eventually update $total_j[k]$ with a value strictly higher than $\stotal_i^{m_i}$. This update, together with the monotonicity of $total_j[\cdot]$ (Proposition~\ref{total:increasing} of Lemma~\ref{lemma:total:ordering}), implies that after some point:

\begin{equation}\forall p_k \in N_\mathcal{G}(p_i): total_j[k] > \stotal_i^{m_i}\end{equation}

\noindent and that $m_i$ is selected in $T_1^j$. We now show by contradiction that $m_i$ eventually progresses to $T_2^j$. Let us assume $m_i$ never meets $T_2^j$'s condition. This means that every time $T_2^j$ is evaluated we have:

\begin{equation}\begin{array}{l}\exists p_{k} \in N_\mathcal{G}(p_i), \exists \big\langle m_k, \langle \scausal_k^{m_k}[\cdot], \stotal_k^{m_k}, k \rangle \big\rangle
\in  pending_j: \hfill{}\\\hfill{} \langle \stotal_k^{m_k}, k\rangle \leq \langle  \stotal_i^m, i \rangle\end{array}\label{eqn:term:T2:never:satisfied}\end{equation}

Note that there could be different $p_k$ and $m_k$ satisfying (\ref{eqn:term:T2:never:satisfied}) in each loop of Task $T$. However, because $N_\mathcal{G}(p_i)$ is finite, the number of timestamps $\langle \stotal_k^{m_k}, k\rangle$ such that $\langle \stotal_k^{m_k}, k\rangle \leq \langle \stotal_i^m, i \rangle$ is also finite. There is therefore one process $p_{k_0}$ and one message $m_{k_0}$ that appear infinitely often in the sequence of $(p_k, m_k)$ that satisfy (\ref{eqn:term:T2:never:satisfied}). Since $m_{k_0}$ can only be inserted once into $pending_j$, this means $m_{k_0}$ remains indefinitely into $T_2^j$, and hence $pending_j$, and is never delivered. (\ref{eqn:term:T2:never:satisfied}) and the fact that $i \neq k_0$ (because $p_i \not\in N_\mathcal{G}(p_i)$) yields

\begin{equation}
  \langle \stotal_k^{m_{k_0}}, k_0\rangle < \langle \stotal_i^m, i \rangle
\end{equation}

\noindent which contradicts our assumption that $m_i$ has the smallest total order timestamps $\langle \stotal_i^{m_i}, i \rangle$ among all messages never delivered to $p_j$. We conclude that after some point $m_i$ remains indefinitely into $T_2^j$.

\begin{equation}m_i \in T_2^j\end{equation}

If we now assume $m_i$ is never returned by $\argmin$ at line~\ref{line:argmin}, we can repeat a similar argument on the finite number of timestamps smaller than $\langle\stotal_i^m, i \rangle$, and the fact that once they have been removed form $pending_j$ (line~\ref{line:remove:from:pending}), messages are never inserted back, and find another message $m_k$ with a strictly smaller time-stamp that $p_j$ that is never delivered. The existence of $m_k$ contradicts again our assumption on the minimality of $m_i$'s timestamp $\langle \stotal_i^m, i \rangle$ among undelivered messages.

This shows that $m_i$ is eventually delivered, and ends our proof by contradiction.
\end{proofT}

\subsubsection*{Causal Order}

We prove the causal order property by induction on the causal order relation $\leadsto_M$.

\begin{lemma}\label{lemma:m1:m2:bcast:by:Pi}
Consider $m_1$ and $m_2$, two messages toco-broadcast by Process $p_i$, with $m_1$ toco-broadcast before $m_2$. If a process $p_j$ toco-delivers $m_2$, then it must have toco-delivered $m_1$ before $m_2$.
\end{lemma}

\begin{proofL}\renewcommand{\toto}{lemma:m1:m2:bcast:by:Pi}
We first consider the order in which the messages were inserted into $pending_j$ (along with their causal timestamps $\scausal_i^{m_{1|2}}$). For $i=j$, $m_1$ was inserted before $m_2$ at line~\ref{line:inserting:sent:m:into:pending} by assumption. For $i\neq j$, we note that if $p_j$ delivers $m_2$ at line~\ref{line:deliver:tocobc}, then $m_2$ was received from $p_i$ at line~\ref{line:tocobc-receive} at some earlier point. Because channels are FIFO, this also means
\begin{equation}m_1\textrm{ was received and added to }pending_j\textrm{ before }m_2\textrm{ was.}\label{equ:m1receivedb4m2}\end{equation}

 We now want to show that when $m_2$ is delivered by $p_j$, $m_1$ is no longer in $pending_j$, which will show that $m_1$ has been delivered before $m_2$. We use an argument by contradiction. Let us assume that
 \begin{equation}\big\langle m_1,\langle \scausal_i^{m_1}, \stotal_i^{m_1}, i \rangle\big\rangle \in pending_j\label{equ:m1withm2inpending}\end{equation}
 \noindent at the start of the iteration of Task $T$ which delivers $m_2$ to $p_j$. From Proposition~\ref{causalii:tocobc:strict:increasing} of Lemma~\ref{lemma:causal:increasing}, we have
\begin{equation}\scausal_i^{m_1} < \scausal_i^{m_2}\label{equ:m1:b4:m2}\end{equation}
\noindent which implies that $m_1$ is selected into $C$ along with $m_2$ (line~\ref{line:C:filtering}):

\[ \big\langle m_1,\langle \scausal_i^{m_1}, \stotal_i^{m_1}, i \rangle\big\rangle \in C\]

\noindent Similarly, from Proposition~\ref{totalii:tobobc:strict:increasing} of Lemma~\ref{lemma:total:ordering} we have:
\begin{equation}\stotal_i^{m_1} < \stotal_i^{m_2}\label{equ:m1:b4:m2:total}\end{equation}

\noindent which implies that $m_1$ must also belong to $T_1$ and $T_2$ (lines~\ref{line:T1:filtering} and~\ref{line:T2:filtering}). (\ref{equ:m1:b4:m2:total}) further implies that $\langle \stotal_i^{m_2}, i \rangle$ is not the minimal $\stotal$ timestamp of $T_2$, and therefore $m_0\neq m_2$ in this iteration of Task $T$. This contradicts our assumption that $m_2$ was delivered in this iteration; shows that (\ref{equ:m1withm2inpending}) must be false; and therefore with (\ref{equ:m1receivedb4m2}) that $m_1$ was delivered before $m_2$.
\end{proofL}

\begin{lemma}\label{lemma:m2:bcst:after:m1:received}
Consider $m_1$ and $m_2$ so that $m_1$ was toco-delivered by a process $p_i$ before $p_i$ toco-broadcasts \(m_2\). If a process $p_j$ toco-delivers $m_2$, then it must have toco-delivered $m_1$ before $m_2$.
\end{lemma}

\begin{proofL}\renewcommand{\toto}{lemma:m2:bcst:after:m1:received}
Let us note $p_k$ the process that has toco-broadcast $m_1$. Because $m_2$ is toco-broadcasts by $p_i$ after $p_i$ toco-delivers $m_1$ and increments $causal_i[k]$ at line~\ref{line:inc:causal:deliver}, we have, using Lemma~\ref{lemma:scausal:link:to:delivery} and Proposition~\ref{causal:increasing} of Lemma~\ref{lemma:causal:increasing}:
\begin{equation}\scausal_k^{m_1}[k] < \scausal_i^{m_2}[k]\label{eqn:scausm1:lt:scausm2}\end{equation}

Because of the condition on set $C$ at line~\ref{line:C:filtering}, when $p_j$ toco-delivers $m_2$ at line~\ref{line:deliver:tocobc}, we further have
\begin{equation}\scausal_i^{m_2}[\cdot] \leq causal_j[\cdot]\label{eqn:scausm2:lt:causj}\end{equation}
and hence using (\ref{eqn:scausm1:lt:scausm2})
\begin{equation}\scausal_k^{m_1}[k] < \scausal_i^{m_2}[k] \leq causal_j[k]\label{eqn:scausm1:lt:causj}\end{equation}

Applying Lemma~\ref{lemma:scausal:link:to:delivery} to (\ref{eqn:scausm1:lt:causj}), we conclude that $p_j$must have toco-delivered $m_1$ when it delivers $m_2$.
\end{proofL}

\begin{theorem}\label{theo:causal:order}
Algorithm~\ref{alg:broadcast-prot} respects causal order.
\end{theorem}

\begin{proofT}\renewcommand{\toto}{theo:causal:order}
We finish the proof by induction on $\leadsto_M$. Let's consider three messages $m_1$, $m_2$, $m_3$ such that
\begin{equation}
  m_1 \leadsto_M m_3 \leadsto_M m_2
\end{equation}
\noindent and such that:
\begin{itemize}
\item if a process toco-delivers $m_3$, it must have toco-delivered $m_1$;
\item if a process toco-delivers $m_2$, it must have toco-delivered $m_3$;
\end{itemize}

We want to show that if a process toco-delivers $m_2$, it must have tolo-delivered $m_1$. This follows from the transitivity of temporal order.
This result together with Lemmas~\ref{lemma:m1:m2:bcast:by:Pi} and~\ref{lemma:m2:bcst:after:m1:received} concludes the proof.
\end{proofT}




\subsubsection*{$\mathcal{G}$-delivery order}

\begin{theorem}\label{theo:gdelivery:order}
Algorithm~\ref{alg:broadcast-prot} respects $\mathcal{G}$-delivery order.
\end{theorem}

\begin{proofT}\renewcommand{\toto}{theo:gdelivery:order}
  Let's consider four processes $p_l$, $p_h$, $p_i$, and $p_j$. $p_l$ and $p_h$ are connected in $\mathcal{G}$. $p_l$ has toco-broadcast a message $m_l$, and $p_h$ has toco-broadcast a message $m_h$. $p_i$ has toco-delivered $m_l$ before $m_h$. $p_j$ has toco-delivered $m_h$. We want to show that $p_j$ has toco-delivered $m_l$ before $m_h$.

We first show that:
\begin{equation}\label{eq:order:on:stot}
 \langle \stotal_h^{m_h}, h \rangle > \langle \stotal_l^{m_l}, l \rangle
\end{equation}
We do so by considering the iteration of the background task $T$ (lines~\ref{line:background:task:start}-\ref{line:background:task:start}) of $p_i$ that toco-delivers $m_l$. Because $p_h \in N_\mathcal{G}(p_l)$, we have
\begin{equation}\label{eq:total:gt:stotl}
  \langle total_i[h],h \rangle > \langle \stotal_l^{m_l}, l \rangle
\end{equation}
\noindent at line~\ref{line:T1:filtering}.

If $m_h$ has not been received by $p_i$ yet, then because of Lemma~\ref{total:increasing}.\ref{totalii:tobobc:strict:increasing}, and because communication channels are FIFO and reliable, we have:
\begin{equation}
   \langle \stotal_h^{m_h}, l \rangle > \langle total_i[h],h \rangle
\end{equation}
\noindent which with (\ref{eq:total:gt:stotl}) yields (\ref{eq:order:on:stot}).

If $m_h$ has already been received by $p_i$, by assumption it has not been toco-delivered yet, and is therefore in $pending_i$. More precisely we have:
\begin{equation}
  \big \langle m_h, \langle \scausal_h^{m_h}[\cdot], \stotal_h^{m_h}, h \rangle \big\rangle \in pending_i
\end{equation}
\noindent which, with $p_h \in N_\mathcal{G}(p_l)$, and the fact that $m_l$ is selected in $T_2^i$ at  line~\ref{line:T2:filtering} also gives us (\ref{eq:order:on:stot}).

We now want to show that $p_j$ must have toco-delivered $m_l$ before $m_h$. The reasoning is somewhat the symmetric of what we have done. We consider the iteration of the background task $T$ of $p_j$ that toco-delivers $m_h$. By the same reasoning as above we have
\begin{equation}\label{eq:total:gt:stoth}
  \langle total_j[l],l \rangle > \langle \stotal_h^{m_h}, h \rangle
\end{equation}
\noindent at line~\ref{line:T1:filtering}.

Because of Lemma~\ref{total:increasing}.\ref{totalii:tobobc:strict:increasing}, and because communication channels are FIFO and reliable, (\ref{eq:total:gt:stoth}) and (\ref{eq:order:on:stot}) imply that $m_l$ has already been received by $p_j$. Because $m_h$ is selected in $T_2^j$ at  line~\ref{line:T2:filtering}, (\ref{eq:order:on:stot}) implies that $m_h$ is no longer in $pending_j$, and so must have been toco-delivered by $p_j$ earlier, which concludes the proof.
\end{proofT}

\algoimplementscccbroadcast*

\begin{proofT}
  \renewcommand{\toto}{theo:proof-toco-bcast}
  \begin{itemize}
  \item Validity and Integrity follow from the integrity and validity of the underlying communication channels, and from how a message $m_j$ is only inserted once into $pending_i$ (at line~\ref{line:inserting:sent:m:into:pending} if $i=j$, at line~\ref{line:inser:mj:pendingi} otherwise) and always removed from $pending_i$ at line~\ref{line:remove:from:pending} before it is toco-delivered by $p_i$ at line~\ref{line:deliver:tocobc};
  \item $\mathcal{G}$-delivery order follows from Theorem~\ref{theo:gdelivery:order};
  \item Causal order follows from Theorem~\ref{theo:causal:order};
  \item Termination follows from Theorem~\ref{theo:termination}.
  \end{itemize}
\end{proofT}
}

\ifdefined\proofInAppendix
The proof, provided in the appendix, combines elements of the proofs of the traditional causal-order~\cite{BSS1991,RST91} and total-order broadcast algorithms~\cite{L78,AW94} on which Algorithm~\ref{alg:broadcast-prot} is based. It relies in particular on the monoticity of the clocks $causal_i[1..n]$ and $total_i[1..n]$, and the reliability and FIFO properties of the underlying communication channels.
\else
  \subsection{Proof of Theorem~\ref{theo:proof-toco-bcast}}
The proof combines elements of the proofs of the traditional causal-order~\cite{BSS1991,RST91} and total-order broadcast algorithms~\cite{L78,AW94} on which Algorithm~\ref{alg:broadcast-prot} is based. It relies in particular on the monoticity of the clocks $causal_i[1..n]$ and $total_i[1..n]$, and the reliability and FIFO properties of the underlying communication channels. We first prove some useful lemmata, before proving termination, causal order, and $\mathcal{G}$-delivery order in intermediate theorems. We finally combine these intermediate results to prove Theorem~\ref{theo:proof-toco-bcast}.

  \proofTocoBcast
\fi


\section{An Algorithm Implementing $\mathcal{G}$-Fisheye (SC,CC)-Consistency}
\label{sec:fisheye-algorithm}

\subsection{The high level object operations read and write}

Algorithm~\ref{alg:SC:CC:consistency-prot} uses the $\mathcal{G}$-fisheye (SC,CC)-broadcast we have just presented to realized $\mathcal{G}$-fisheye (SC,CC)-consistency using a fast-read strategy. This algorithm is derived from the fast-read algorithm for sequential consistency proposed by Attiya and Welch \cite{AW94}, in which the total order broadcast has been replaced by our $\mathcal{G}$-fisheye (SC,CC)-broadcast.

\begin{algorithm}
\caption{Implementing $\mathcal{G}$-fisheye (SC,CC)-consistency
, executed by $p_i$}
\label{alg:SC:CC:consistency-prot}
\begin{algorithmic}[1]
  \setlength{\arraycolsep}{2pt}
  \Function{$X$.write}{$v$}\label{line:start:write}
    \State \Call{TOCO\_broadcast}{\textsc{write}$(X,v,i)$}\label{line:tocobc:write}
    \State $delivered_i \leftarrow false$ ;
    \State \textbf{wait until} $delivered_i=true$\label{line:pi:waits}
  \EndFunction\label{line:end:write}

  \skipSomeSpace

  \Function{$X$.read}{}
    \State \textbf{return} $v_x$
  \EndFunction

  \skipSomeSpace

  \OnTocoDeliver{\textsc{write}$(X,v,j)$}
    \State $v_x \leftarrow v$ ; \label{line:allocating:v:to:vx}
    \State \textbf{if} $(i=j)$ \textbf{then} $delivered_i \leftarrow true$ \textbf{endif}\label{line:toggle:delivered}
  \EndTocoDeliver

\end{algorithmic}
\end{algorithm}

The \func{write}$(X,v)$ operation uses the $\mathcal{G}$-fisheye
(SC,CC)-broadcast to propagate the new value of the variable $X$. To
insure any other write operations that must be seen \emph{before} \func{write}$(X,v)$ by $p_i$ are properly processed, $p_i$ enters a waiting loop (line~\ref{line:pi:waits}), which ends after the message \textsc{write}$(X,v,i)$ that has been toco-broadcast at line~\ref{line:tocobc:write} is toco-delivered at line~\ref{line:toggle:delivered}.

The \func{read}$(X)$ operation simply returns the local copy $v_x$ of $X$. These local copies are updated in the background when \textsc{write}$(X,v,j)$ messages are toco-delivered.

\begin{restatable}{theorem}{algoimplemeentscccconst}
\label{theo:proof-that-algorithm}
  Algorithm~\ref{alg:SC:CC:consistency-prot} implements $\mathcal{G}$-fisheye (SC,CC)-consistency.
\end{restatable}


\newcommand{\proofSCCConsistency}{


\newcommand{\opwrite}{w}
\newcommand{\opread}{r}
\newcommand{\rfrel}{\overset{\mathit{rf}}{\rightarrow}}
\newcommand{\porel}{\overset{\mathit{po}}{\rightarrow}}

For readability, we denote in the following $\opread_p(X,v)$ the read operation invoked by process $p$ on object $X$ that returns a value $v$ ($X.\text{read} \rightarrow v$), and $\opwrite_p(X,v)$ the write operation of value $v$ on object $X$ invoked by process $p$ ($X.\text{write}(v)$). We may omit the name of the process when not needed.

Let us consider a history $\widehat{H}=(H,\porel_H)$ that captures an execution of Algorithm~\ref{alg:SC:CC:consistency-prot}, i.e., $\porel_H$ captures the sequence of operations in each process (process order, $po$ for short). We construct the causal order $\leadsto_H$ required by the definition of Section~\ref{sec:fish-cons-pair} in the following, classical, manner:
\begin{itemize}
\item We connect each read operation $\opread_p(X,v)=X.\text{read} \rightarrow v$ invoked by process $p$ (with $v\neq \bot$, the initial value) to the write operation $\opwrite(X,v)=X.\text{write}(v)$ that generated the \textsc{write}$(X,v)$ message carrying the value $v$ to $p$ (line~\ref{line:allocating:v:to:vx} in Algorithm~\ref{alg:SC:CC:consistency-prot}). In other words, we add an edge $\langle\opwrite(X,v) \overset{\mathit{rf}}{\rightarrow} \opread_p(X,v)\rangle$ to $\porel_H$ (with $\opwrite$ and $\opread_p$ as described above) for each read operation $\opread_p(X,v) \in H$ that does not return the initial value $\bot$. We connect initial read operations $\opread(X,\bot)$ to an $\bot$ element that we add to $H$.

We call these additional relations \emph{read-from links} (noted $\rfrel$).
\item We take $\leadsto_H$ to be the transitive closure of the resulting relation.
\end{itemize}

$\leadsto_H$ is acyclic, as assuming otherwise would imply at least one of the \textsc{write}$(X,v)$ messages was received before it was sent. $\leadsto_H$ is therefore an order. We now need to show $\leadsto_H$ is a causal order in the sense of the definition of Section~\ref{sec:causal-consistency}, i.e., that the result of each read operation $\opread(X,v)$ is the value of the latest write $\opwrite(X,v)$ that occurred before $\opread(X,v)$ in $\leadsto_H$ (said differently, that no read returns an overwritten value).

\begin{lemma}\label{lemma:leadstoH:causal}
  $\leadsto_H$ is a causal order.
\end{lemma}

\begin{proofL}\renewcommand{\toto}{lemma:leadstoH:causal}
We show this by contradiction. We assume without loss of generality that all values written are distinct. Let us consider $\opwrite_p(X,v)$ and $\opread_q(X,v)$ so that $\opwrite_p(X,v) \rfrel \opread_q(X,v)$, which implies $\opwrite_p(X,v) \leadsto_H \opread_q(X,v)$. Let us assume there exists a second write operation $\opwrite_r(X,v')\neq \opwrite_p(X,v)$ on the same object, so that
\begin{equation}
 \opwrite_p(X,v) \leadsto_H \opwrite_r(X,v') \leadsto_H \opread_q(X,v)
\end{equation}
\noindent (illustrated in Figure~\ref{fig:leadstoHcausal}).
$\opwrite_p(X,v) \leadsto_H \opwrite_r(X,v')$ means we can find a sequence of operations $op_i \in H$ so that
\begin{equation}
  \opwrite_p(X,v) \rightarrow_0 op_0 ... \rightarrow_i op_i \rightarrow_{i+1} ... \rightarrow_k \opwrite_r(X,v')
\end{equation}
\noindent with $\rightarrow_i \in \{\porel_H,\rfrel\}, \forall i\in [1,k]$. The semantics of $\porel_H$ and $\rfrel$ means we can construct a sequence of causally related (SC,CC)-broadcast messages $m_i \in M$ between the messages that are toco-broadcast by the operations $\opwrite_p(X,v)$ and $\opwrite_r(X,v')$, which we note $\textsc{write}_p(X,v)$ and $\textsc{write}_r(X,v')$ respectively:
\begin{equation}
  \textsc{write}_p(X,v) = m_0 \leadsto_M m_1 ... \leadsto_M m_i \leadsto_M ... \leadsto_M m_{k'} = \textsc{write}_r(X,v')
\end{equation}
\noindent where $\leadsto_M$ is the message causal order introduced in Section~\ref{sec:mathcalg-fisheye-sc}. We conclude that $\textsc{write}_p(X,v) \leadsto_M \textsc{write}_r(X,v')$, i.e., that $\textsc{write}_p(X,v)$ belongs to the causal past of $\textsc{write}_r(X,v')$, and hence that $q$ in Figure~\ref{fig:leadstoHcausal} toco-delivers $\textsc{write}_r(X,v')$ after $\textsc{write}_p(X,v)$.


\begin{figure}[tbh]
\centering
\usetikzlibrary{calc}
\begin{tikzpicture}[node distance=0.9cm]
  \renewcommand{\makeoplabel}[1]{}
  \renewcommand{\writesyntax}[2]{$w_{\processid}(#1,#2)$}
  \renewcommand{\readsyntax}[2]{$r_{\processid}(#1,#2)$}

  \ProcessWithId{p}{$p$}{6}{%
    \writeOp{X}{v}{0.5}{2.0}{}
  }
  \ProcessWithId{q}{$q$}{6}{%
    \readOp{X}{v}{4.5}{6.0}{}
  }
  \ProcessWithId{r}{$r$}{6}{%
    \writeOp{X}{v'}{2.5}{4.0}{}
  }

  \procmessage{p}{1.25}{q}{5.25}{$\rfrel$}{bend right=2}
  \procmessage{p}{1.25}{r}{2.5}{$\leadsto_H$}{bend right=0,dashed}
  \procmessage{r}{4   }{q}{5.25}{$\leadsto_H$}{bend right=2,dashed}

\end{tikzpicture}
\caption{Proving that $\leadsto_H$ is causal by contradiction}
\label{fig:leadstoHcausal}
\end{figure}

We now want to show that $\textsc{write}_r(X,v')$ is toco-delivered by $q$ before $q$ executes $\opread_q(X,v)$. We can apply the same reasoning as above to $\opwrite_r(X,v') \leadsto_H \opread_q(X,v)$, yielding another sequence of operations $op'_i \in H$:
\begin{equation}
  \opwrite_r(X,v') \rightarrow'_0 op'_0 ... \rightarrow'_i op'_i \rightarrow'_{i+1} ... \rightarrow'_{k''} \opread_q(X,v)
\end{equation}
\noindent with $\rightarrow'_i \in \{\porel_H,\rfrel\}$. Because $\opread_q(X,v)$ does not generate any (SC,CC)-broadcast message, we need to distinguish the case where all $op'_i$ relations correspond to the process order $\porel_H$ (i.e., $op'_i = \porel_H, \forall i$). In this case, $r=q$, and the blocking behavior of $X$.\func{write}() (line~\ref{line:pi:waits} of Algorithm~\ref{alg:SC:CC:consistency-prot}), insures that $\textsc{write}_r(X,v')$ is toco-delivered by $q$ before executing $\opread_q(X,v)$. If at least one $op'_i$ corresponds to the read-from relation, we can consider the latest one in the sequence, which will denote the toco-delivery of a $\textsc{write}_z(Y,w)$ message by $q$, with $\textsc{write}_r(X,v') \leadsto_M \textsc{write}_z(Y,w)$. From the causality of the (SC,CC)-broadcast, we also conclude that $\textsc{write}_r(X,v')$ is toco-delivered by $q$ before executing $\opread_q(X,v)$.

Because q toco-delivers $\textsc{write}_p(X,v)$ before $\textsc{write}_r(X,v')$, and toco-delivers $\textsc{write}_r(X,v')$ before it executes $\opread_q(X,v)$, we conclude that the value $v$ of $v_x$ is overwritten by $v'$ at line~\ref{line:allocating:v:to:vx} of Algorithm~\ref{alg:SC:CC:consistency-prot}, and that $\opread_q(X,v)$ does not return $v$, contradicting our assumption that $\opwrite_p(X,v) \rfrel \opread_q(X,v)$, and concluding our proof that $\leadsto_H$ is a causal order.
\end{proofL}

\newcommand{\leadmoon}{\overset{\leftmoon}{\leadsto}}
\newcommand{\leadsmallstar}{\overset{\medstar}{\leadsto}}

\newcommand{\leadmooninc}[1]{\leadmoon_H^{\raisebox{-0.3em}{$\scriptstyle #1$}}}
\newcommand{\leadsmallstarinc}[1]{\leadsmallstar_H^{\raisebox{-0.3em}{$\scriptstyle #1$}}}
\newcommand{\leadstostarinc}[1]{\leadstostar_H^{\raisebox{-0.3em}{$\scriptstyle #1$}}}

\newcommand{\tocorel}{\overset{ww}{\rightarrow}}

To construct $\leadstostar_{H,\mathcal{G}}$, as required by the definition of (SC,CC)-consistency (Section~\ref{sec:fish-cons-pair}), we need to order the write operations of neighboring processes in the proximity graph $\mathcal{G}$. We do so as follows:
\begin{itemize}
\item
We add an edge $\opwrite_p(X,v) \tocorel \opwrite_q(Y,w)$ to $\leadsto_H$ for each pair of write operations $\opwrite_p(X,v)$ and $\opwrite_q(Y,w)$ in $H$ such that:
\begin{itemize}
\item $(p,q)\in E_{\mathcal{G}}$ (i.e., $p$ and $q$ are connected in $\mathcal{G}$);
\item $\opwrite_p(X,v)$ and $\opwrite_q(Y,w)$ are not ordered in $\leadsto_H$;
\item The broadcast message $\textsc{write}_p(X,v)$ of $\opwrite_p(X,v)$ has been toco-delivered before the broadcast message $\textsc{write}_p(Y,w)$ of $\opwrite_q(Y,w)$ by all processes.
\end{itemize}

We call these additional edges \emph{ww links} (noted $\tocorel$).
\item We take $\leadstostar_{H,\mathcal{G}}$ to be the recursive closure of the relation we obtain.
\end{itemize}

$\leadstostar_{H,\mathcal{G}}$ is acyclic, as assuming otherwise would imply that the underlying (SC,CC)-broadcast violates causality. Because of the $\mathcal{G}$-delivery order and termination of the toco-broadcast (Section~\ref{sec:mathcalg-fisheye-sc}), we know all pairs of $\textsc{write}_p(X,v)$ and $\textsc{write}_p(Y,w)$ messages with $(p,q)\in E_{\mathcal{G}}$ as defined above are toco-delivered in the same order by all processes. This insures that all write operations of neighboring processes in $\mathcal{G}$ are ordered in $\leadstostar_{H,\mathcal{G}}$.

We need to show that $\leadstostar_{H,\mathcal{G}}$ remains a causal order, i.e., that no read in $\leadstostar_{H,\mathcal{G}}$ returns an overwritten value.

\begin{lemma}\label{lemma:leadmoon:causal}
  $\leadstostar_{H,\mathcal{G}}$ is a causal order.
\end{lemma}

\begin{proofL}\renewcommand{\toto}{lemma:leadmoon:causal}
We extend the original causal order  $\leadsto_M$ on the messages of an (SC,CC)-broadcast execution with the following order $\leadsto_M^{\mathcal{G}}$:

\noindent $m_1 \leadsto_M^{\mathcal{G}} m_2$ if
\begin{itemize}
\item $m_1 \leadsto_M m_2$; or
\item $m_1$ was sent by $p$, $m_2$ by $q$, $(p,q)\in E_{\mathcal{G}}$, and $m_1$ is toco-delivered before $m_2$ by all processes; or
\item there exists a message $m_3$ so that $m_1 \leadsto_M^{\mathcal{G}} m_3$ and $m_3 \leadsto_M^{\mathcal{G}} m_2$.
\end{itemize}
$\leadsto_M^{\mathcal{G}}$ captures the order imposed by an execution of an (SC,CC)-broadcast on its messages. The proof is then identical to that of Lemma~\ref{lemma:leadstoH:causal}, except that we use the order $\leadsto_M^{\mathcal{G}}$, instead of $\leadsto_M$.
\end{proofL}

\newcommand{\rwrel}{\overset{rw}{\rightarrow}}

\algoimplemeentscccconst*


\begin{proofT}\renewcommand{\toto}{theo:proof-that-algorithm}
The order $\leadstostar_{H,\mathcal{G}}$ we have just constructed fulfills the conditions required by the definition of $\mathcal{G}$-fisheye (SC,CC)-consistency (Section~\ref{sec:fish-cons-pair}):
\begin{itemize}
\item by construction $\leadstostar_{H,\mathcal{G}}$ subsumes $\leadsto_H$ ($\leadsto_H\;\subseteq\; \leadstostar_{H,\mathcal{G}}$);
\item also by construction $\leadstostar_{H,\mathcal{G}}$, any pair of write operations invoked by processes $p$,$q$ that are neighbors in $\mathcal{G}$ are ordered in $\leadstostar_{H,\mathcal{G}}$; i.e., $\mbox{$(\leadstostar_{H,\mathcal{G}})$}|(\{p,q\}\cap W)$ is a total order.
\end{itemize}

To finish the proof, we choose, for each process $p_i$, $\widehat{S}_i$ as one of the topological sorts of $\mbox{$(\leadstostar_{H,\mathcal{G}})$}|(p_i+W)$, following the approach of \cite{MRZ1995,R13-Dist-alg}. $\widehat{S}_i$ is sequential by construction. Because $\leadstostar_{H,\mathcal{G}}$ is causal, $\widehat{S}_i$ is legal. Because $\leadstostar_{H,\mathcal{G}}$ respects $\porel_H$, $\widehat{S}_i$ is equivalent to $\widehat{H}{|(p_i+W)}$. Finally, $\widehat{S}_i$ respects $\mbox{$(\leadstostar_{H,\mathcal{G}})$}|(p_i+W)$ by construction.
\end{proofT}






}

\ifdefined\proofInAppendix
The proof (detailed in the appendix) uses the causal order on messages $\leadsto_M$ provided by the $\mathcal{G}$-fisheye (SC,CC)-broadcast to construct the causal order on operations $\leadsto_H$. It then gradually extends $\leadsto_H$ to obtain $\leadstostar_{H,\mathcal{G}}$. It first uses the property of the broadcast algorithm on messages toco-broadcast by processes that are neighbors in $\mathcal{G}$, and then adapts the technique used in~\cite{MRZ1995,R13-Dist-alg} to show that WW (write-write) histories are sequentially consistent. The individual histories $\widehat{S}_i$ are obtained by taking a topological sort of $(\leadstostar_{H,\mathcal{G}}){|(p_i+W)}$.
\else
  \subsection{Proof of Theorem~\ref{theo:proof-that-algorithm}}
The proof uses the causal order on messages $\leadsto_M$ provided by the $\mathcal{G}$-fisheye (SC,CC)-broadcast to construct the causal order on operations $\leadsto_H$. It then gradually extends $\leadsto_H$ to obtain $\leadstostar_{H,\mathcal{G}}$. It first uses the property of the broadcast algorithm on messages to-broadcast by processes that are neighbors in $\mathcal{G}$, and then adapts the technique used in~\cite{MRZ1995,R13-Dist-alg} to show that WW (write-write) histories are sequentially consistent. The individual histories $\widehat{S}_i$ are obtained by taking a topological sort of $(\leadstostar_{H,\mathcal{G}}){|(p_i+W)}$.



\newcommand{\opwrite}{w}
\newcommand{\opread}{r}
\newcommand{\rfrel}{\overset{\mathit{rf}}{\rightarrow}}
\newcommand{\porel}{\overset{\mathit{po}}{\rightarrow}}

For readability, we denote in the following $\opread_p(X,v)$ the read operation invoked by process $p$ on object $X$ that returns a value $v$ ($X.\text{read} \rightarrow v$), and $\opwrite_p(X,v)$ the write operation of value $v$ on object $X$ invoked by process $p$ ($X.\text{write}(v)$). We may omit the name of the process when not needed.

Let us consider a history $\widehat{H}=(H,\porel_H)$ that captures an execution of Algorithm~\ref{alg:SC:CC:consistency-prot}, i.e., $\porel_H$ captures the sequence of operations in each process (process order, $po$ for short). We construct the causal order $\leadsto_H$ required by the definition of Section~\ref{sec:fish-cons-pair} in the following, classical, manner:
\begin{itemize}
\item We connect each read operation $\opread_p(X,v)=X.\text{read} \rightarrow v$ invoked by process $p$ (with $v\neq \bot$, the initial value) to the write operation $\opwrite(X,v)=X.\text{write}(v)$ that generated the \textsc{write}$(X,v)$ message carrying the value $v$ to $p$ (line~\ref{line:allocating:v:to:vx} in Algorithm~\ref{alg:SC:CC:consistency-prot}). In other words, we add an edge $\langle\opwrite(X,v) \overset{\mathit{rf}}{\rightarrow} \opread_p(X,v)\rangle$ to $\porel_H$ (with $\opwrite$ and $\opread_p$ as described above) for each read operation $\opread_p(X,v) \in H$ that does not return the initial value $\bot$. We connect initial read operations $\opread(X,\bot)$ to an $\bot$ element that we add to $H$.

We call these additional relations \emph{read-from links} (noted $\rfrel$).
\item We take $\leadsto_H$ to be the transitive closure of the resulting relation.
\end{itemize}

$\leadsto_H$ is acyclic, as assuming otherwise would imply at least one of the \textsc{write}$(X,v)$ messages was received before it was sent. $\leadsto_H$ is therefore an order. We now need to show $\leadsto_H$ is a causal order in the sense of the definition of Section~\ref{sec:causal-consistency}, i.e., that the result of each read operation $\opread(X,v)$ is the value of the latest write $\opwrite(X,v)$ that occurred before $\opread(X,v)$ in $\leadsto_H$ (said differently, that no read returns an overwritten value).

\begin{lemma}\label{lemma:leadstoH:causal}
  $\leadsto_H$ is a causal order.
\end{lemma}

\begin{proofL}\renewcommand{\toto}{lemma:leadstoH:causal}
We show this by contradiction. We assume without loss of generality that all values written are distinct. Let us consider $\opwrite_p(X,v)$ and $\opread_q(X,v)$ so that $\opwrite_p(X,v) \rfrel \opread_q(X,v)$, which implies $\opwrite_p(X,v) \leadsto_H \opread_q(X,v)$. Let us assume there exists a second write operation $\opwrite_r(X,v')\neq \opwrite_p(X,v)$ on the same object, so that
\begin{equation}
 \opwrite_p(X,v) \leadsto_H \opwrite_r(X,v') \leadsto_H \opread_q(X,v)
\end{equation}
\noindent (illustrated in Figure~\ref{fig:leadstoHcausal}).
$\opwrite_p(X,v) \leadsto_H \opwrite_r(X,v')$ means we can find a sequence of operations $op_i \in H$ so that
\begin{equation}
  \opwrite_p(X,v) \rightarrow_0 op_0 ... \rightarrow_i op_i \rightarrow_{i+1} ... \rightarrow_k \opwrite_r(X,v')
\end{equation}
\noindent with $\rightarrow_i \in \{\porel_H,\rfrel\}, \forall i\in [1,k]$. The semantics of $\porel_H$ and $\rfrel$ means we can construct a sequence of causally related (SC,CC)-broadcast messages $m_i \in M$ between the messages that are toco-broadcast by the operations $\opwrite_p(X,v)$ and $\opwrite_r(X,v')$, which we note $\textsc{write}_p(X,v)$ and $\textsc{write}_r(X,v')$ respectively:
\begin{equation}
  \textsc{write}_p(X,v) = m_0 \leadsto_M m_1 ... \leadsto_M m_i \leadsto_M ... \leadsto_M m_{k'} = \textsc{write}_r(X,v')
\end{equation}
\noindent where $\leadsto_M$ is the message causal order introduced in Section~\ref{sec:mathcalg-fisheye-sc}. We conclude that $\textsc{write}_p(X,v) \leadsto_M \textsc{write}_r(X,v')$, i.e., that $\textsc{write}_p(X,v)$ belongs to the causal past of $\textsc{write}_r(X,v')$, and hence that $q$ in Figure~\ref{fig:leadstoHcausal} toco-delivers $\textsc{write}_r(X,v')$ after $\textsc{write}_p(X,v)$.


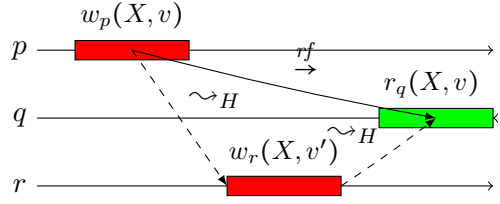
\begin{figure}[tbh]
\centering
\usetikzlibrary{calc}
\begin{tikzpicture}[node distance=0.9cm]
  \renewcommand{\makeoplabel}[1]{}
  \renewcommand{\writesyntax}[2]{$w_{\processid}(#1,#2)$}
  \renewcommand{\readsyntax}[2]{$r_{\processid}(#1,#2)$}

  \ProcessWithId{p}{$p$}{6}{%
    \writeOp{X}{v}{0.5}{2.0}{}
  }
  \ProcessWithId{q}{$q$}{6}{%
    \readOp{X}{v}{4.5}{6.0}{}
  }
  \ProcessWithId{r}{$r$}{6}{%
    \writeOp{X}{v'}{2.5}{4.0}{}
  }

  \procmessage{p}{1.25}{q}{5.25}{$\rfrel$}{bend right=2}
  \procmessage{p}{1.25}{r}{2.5}{$\leadsto_H$}{bend right=0,dashed}
  \procmessage{r}{4   }{q}{5.25}{$\leadsto_H$}{bend right=2,dashed}

\end{tikzpicture}
\caption{Proving that $\leadsto_H$ is causal by contradiction}
\label{fig:leadstoHcausal}
\end{figure}

We now want to show that $\textsc{write}_r(X,v')$ is toco-delivered by $q$ before $q$ executes $\opread_q(X,v)$. We can apply the same reasoning as above to $\opwrite_r(X,v') \leadsto_H \opread_q(X,v)$, yielding another sequence of operations $op'_i \in H$:
\begin{equation}
  \opwrite_r(X,v') \rightarrow'_0 op'_0 ... \rightarrow'_i op'_i \rightarrow'_{i+1} ... \rightarrow'_{k''} \opread_q(X,v)
\end{equation}
\noindent with $\rightarrow'_i \in \{\porel_H,\rfrel\}$. Because $\opread_q(X,v)$ does not generate any (SC,CC)-broadcast message, we need to distinguish the case where all $op'_i$ relations correspond to the process order $\porel_H$ (i.e., $op'_i = \porel_H, \forall i$). In this case, $r=q$, and the blocking behavior of $X$.\func{write}() (line~\ref{line:pi:waits} of Algorithm~\ref{alg:SC:CC:consistency-prot}), insures that $\textsc{write}_r(X,v')$ is toco-delivered by $q$ before executing $\opread_q(X,v)$. If at least one $op'_i$ corresponds to the read-from relation, we can consider the latest one in the sequence, which will denote the toco-delivery of a $\textsc{write}_z(Y,w)$ message by $q$, with $\textsc{write}_r(X,v') \leadsto_M \textsc{write}_z(Y,w)$. From the causality of the (SC,CC)-broadcast, we also conclude that $\textsc{write}_r(X,v')$ is toco-delivered by $q$ before executing $\opread_q(X,v)$.

Because q toco-delivers $\textsc{write}_p(X,v)$ before $\textsc{write}_r(X,v')$, and toco-delivers $\textsc{write}_r(X,v')$ before it executes $\opread_q(X,v)$, we conclude that the value $v$ of $v_x$ is overwritten by $v'$ at line~\ref{line:allocating:v:to:vx} of Algorithm~\ref{alg:SC:CC:consistency-prot}, and that $\opread_q(X,v)$ does not return $v$, contradicting our assumption that $\opwrite_p(X,v) \rfrel \opread_q(X,v)$, and concluding our proof that $\leadsto_H$ is a causal order.
\end{proofL}

\newcommand{\leadmoon}{\overset{\leftmoon}{\leadsto}}
\newcommand{\leadsmallstar}{\overset{\medstar}{\leadsto}}

\newcommand{\leadmooninc}[1]{\leadmoon_H^{\raisebox{-0.3em}{$\scriptstyle #1$}}}
\newcommand{\leadsmallstarinc}[1]{\leadsmallstar_H^{\raisebox{-0.3em}{$\scriptstyle #1$}}}
\newcommand{\leadstostarinc}[1]{\leadstostar_H^{\raisebox{-0.3em}{$\scriptstyle #1$}}}

\newcommand{\tocorel}{\overset{ww}{\rightarrow}}

To construct $\leadstostar_{H,\mathcal{G}}$, as required by the definition of (SC,CC)-consistency (Section~\ref{sec:fish-cons-pair}), we need to order the write operations of neighboring processes in the proximity graph $\mathcal{G}$. We do so as follows:
\begin{itemize}
\item
We add an edge $\opwrite_p(X,v) \tocorel \opwrite_q(Y,w)$ to $\leadsto_H$ for each pair of write operations $\opwrite_p(X,v)$ and $\opwrite_q(Y,w)$ in $H$ such that:
\begin{itemize}
\item $(p,q)\in E_{\mathcal{G}}$ (i.e., $p$ and $q$ are connected in $\mathcal{G}$);
\item $\opwrite_p(X,v)$ and $\opwrite_q(Y,w)$ are not ordered in $\leadsto_H$;
\item The broadcast message $\textsc{write}_p(X,v)$ of $\opwrite_p(X,v)$ has been toco-delivered before the broadcast message $\textsc{write}_p(Y,w)$ of $\opwrite_q(Y,w)$ by all processes.
\end{itemize}

We call these additional edges \emph{ww links} (noted $\tocorel$).
\item We take $\leadstostar_{H,\mathcal{G}}$ to be the recursive closure of the relation we obtain.
\end{itemize}

$\leadstostar_{H,\mathcal{G}}$ is acyclic, as assuming otherwise would imply that the underlying (SC,CC)-broadcast violates causality. Because of the $\mathcal{G}$-delivery order and termination of the toco-broadcast (Section~\ref{sec:mathcalg-fisheye-sc}), we know all pairs of $\textsc{write}_p(X,v)$ and $\textsc{write}_p(Y,w)$ messages with $(p,q)\in E_{\mathcal{G}}$ as defined above are toco-delivered in the same order by all processes. This insures that all write operations of neighboring processes in $\mathcal{G}$ are ordered in $\leadstostar_{H,\mathcal{G}}$.

We need to show that $\leadstostar_{H,\mathcal{G}}$ remains a causal order, i.e., that no read in $\leadstostar_{H,\mathcal{G}}$ returns an overwritten value.

\begin{lemma}\label{lemma:leadmoon:causal}
  $\leadstostar_{H,\mathcal{G}}$ is a causal order.
\end{lemma}

\begin{proofL}\renewcommand{\toto}{lemma:leadmoon:causal}
We extend the original causal order  $\leadsto_M$ on the messages of an (SC,CC)-broadcast execution with the following order $\leadsto_M^{\mathcal{G}}$:

\noindent $m_1 \leadsto_M^{\mathcal{G}} m_2$ if
\begin{itemize}
\item $m_1 \leadsto_M m_2$; or
\item $m_1$ was sent by $p$, $m_2$ by $q$, $(p,q)\in E_{\mathcal{G}}$, and $m_1$ is toco-delivered before $m_2$ by all processes; or
\item there exists a message $m_3$ so that $m_1 \leadsto_M^{\mathcal{G}} m_3$ and $m_3 \leadsto_M^{\mathcal{G}} m_2$.
\end{itemize}
$\leadsto_M^{\mathcal{G}}$ captures the order imposed by an execution of an (SC,CC)-broadcast on its messages. The proof is then identical to that of Lemma~\ref{lemma:leadstoH:causal}, except that we use the order $\leadsto_M^{\mathcal{G}}$, instead of $\leadsto_M$.
\end{proofL}

\newcommand{\rwrel}{\overset{rw}{\rightarrow}}

\algoimplemeentscccconst*


\begin{proofT}\renewcommand{\toto}{theo:proof-that-algorithm}
The order $\leadstostar_{H,\mathcal{G}}$ we have just constructed fulfills the conditions required by the definition of $\mathcal{G}$-fisheye (SC,CC)-consistency (Section~\ref{sec:fish-cons-pair}):
\begin{itemize}
\item by construction $\leadstostar_{H,\mathcal{G}}$ subsumes $\leadsto_H$ ($\leadsto_H\;\subseteq\; \leadstostar_{H,\mathcal{G}}$);
\item also by construction $\leadstostar_{H,\mathcal{G}}$, any pair of write operations invoked by processes $p$,$q$ that are neighbors in $\mathcal{G}$ are ordered in $\leadstostar_{H,\mathcal{G}}$; i.e., $\mbox{$(\leadstostar_{H,\mathcal{G}})$}|(\{p,q\}\cap W)$ is a total order.
\end{itemize}

To finish the proof, we choose, for each process $p_i$, $\widehat{S}_i$ as one of the topological sorts of $\mbox{$(\leadstostar_{H,\mathcal{G}})$}|(p_i+W)$, following the approach of \cite{MRZ1995,R13-Dist-alg}. $\widehat{S}_i$ is sequential by construction. Because $\leadstostar_{H,\mathcal{G}}$ is causal, $\widehat{S}_i$ is legal. Because $\leadstostar_{H,\mathcal{G}}$ respects $\porel_H$, $\widehat{S}_i$ is equivalent to $\widehat{H}{|(p_i+W)}$. Finally, $\widehat{S}_i$ respects $\mbox{$(\leadstostar_{H,\mathcal{G}})$}|(p_i+W)$ by construction.
\end{proofT}







\fi

\section{Conclusion}
\label{sec:conclusion}

This work was motivated by the increasing popularity of geographically
distributed systems. We have presented a framework that enables to formally
define and reason about mixed consistency conditions in which the operations
invoked by nearby processes obey stronger consistency requirements than
operations invoked by remote ones. The framework is based on the concept of a
proximity graph, which captures the ``closeness'' relationship between
processes. As an example, we have formally defined $\mathcal{G}$-fisheye
(SC,CC)-consistency, which combines sequential consistency for operations by
close processes with causal consistency among all operations. We have also
provided a formally proven protocol for implementing $\mathcal{G}$-fisheye
(SC,CC)-consistency.

Another natural example that has been omitted from this paper for brevity
is $\mathcal{G}$-fisheye (LIN,SC)-consistency, which combines
linearizability for operations by nearby nodes with an overall
sequential consistency guarantee.

The significance of our approach is that the definitions of consistency
conditions are functional rather than operational. That is, they are
independent of a specific implementation, and provide a clear rigorous
understanding of the provided semantics. This clear understanding and formal underpinning comes with improved complexity and performance, as illustrated in our implementation of $\mathcal{G}$-fisheye
(SC,CC)-consistency, in which operations can terminate without waiting to
synchronize with remote parts of the system.

More generally, we expect the general philosophy we have presented to extend to Convergent Replicated Datatypes (CRDT) in which not all operations are commutative \cite{OUMI2006}. These CRDTs usually require at a minimum causal communications to implement eventual consistency. The hybridization we have proposed opens up the path of CRDTs which are globally eventually consistent, and locally sequentially consistent, a route we plan to explore in future work.




%
%
%
%


\section*{Acknowledgments}
{
This work has been partially supported by a French government support granted to the CominLabs excellence laboratory (Project \emph{DeSceNt: Plug-based Decentralized Social Network}) and managed by the French National Agency for Research (ANR) in the "Investing for the Future" program under reference Nb. ANR-10-LABX-07-01, and by the SocioPlug Project funded by French National Agency for Research (ANR), under program ANR INFRA (ANRANR-13-INFR-0003).} We would also like to thank Matthieu Perrin for many enlightening discussions on the topic of weak consistency models, and for pointing out a flaw in an earlier definition of fisheye consistency.


\ifdefined\proofInAppendix

\newpage

\section*{Appendix: Detail of Proofs}

\subsection*{Proof that Algorithm~\ref{alg:broadcast-prot} implements a $\mathcal{G}$-fisheye (SC,CC)-broadcast}

\proofTocoBcast

\subsection*{Proof that Algorithm~\ref{alg:SC:CC:consistency-prot} implements $\mathcal{G}$-fisheye (SC,CC)-consistency}

\fi

\end{document}